# A Multi-Scale Finite Element Method for Investigating Fiber Remodeling in Hypertrophic Cardiomyopathy

**Mohammad Mehri[1], Kenneth S. Campbell[2], Lik Chuan Lee[3], Jonathan F. Wenk [1,4*]**

[1]Department of Mechanical and Aerospace Engineering, University of Kentucky, Lexington, Kentucky, USA

[2]Division of Cardiovascular Medicine and Department of Physiology, University of Kentucky, Lexington, KY

[3]Department of Mechanical Engineering, Michigan State University, East Lansing, Michigan, USA

[4]Department of Surgery, University of Kentucky, Lexington, Kentucky, USA

*Corresponding Author:

Jonathan F. Wenk, PhD

University of Kentucky

Department of Mechanical and Aerospace Engineering

269 Ralph G. Anderson Building

Lexington, KY 40506-0503

Phone: (859) 218-0658

Fax: (859) 257-3304

Email: jonathan.wenk@uky.edu

## Abstract:

A significant hallmark of hypertrophic cardiomyopathy (HCM) is fiber disarray, which is associated with various cardiac events such as heart failure. Quantifying fiber disarray remains critical for understanding the disease's complex pathophysiology. This study investigates the role of heterogeneous HCM-induced cellular abnormalities in the development of fiber disarray and their subsequent impact on cardiac pumping function. Fiber disarray is predicted using a stress-based law to reorient myofibers and collagen within a multiscale finite element cardiac modeling framework, MyoFE. Specifically, the model is used to quantify the distinct impacts of heterogeneous distributions of hypercontractility, hypocontractility, and fibrosis on fiber disarray development and examines their effect on functional characteristics of the heart. Our results show that heterogenous cell level abnormalities highly disrupt the normal mechanics of myocardium and lead to significant fiber disarray. The pattern of disarray varies depending on the specific perturbation, offering valuable insights into the progression of HCM. Despite the random distribution of perturbed regions within the cardiac muscle, significantly higher fiber disarray is observed near the epicardium compared to the endocardium across all perturbed left ventricle (LV) models. This regional difference in fiber disarray, irrespective of perturbation severity, aligns with previous DT-MRI studies, highlighting the role of regional myocardial mechanics in the development of fiber disarray. Furthermore, cardiac performance declined in the remodeled LVs, particularly in those with fibrosis and hypocontractility. These findings provide important insights into the structural and functional consequences of HCM and offer a framework for future investigations into therapeutic interventions targeting cardiac remodeling.

## 1. Introduction

Hypertrophic cardiomyopathy (HCM) is the most prevalent inherited heart disease, affecting over 1 in 200 individuals [1, 2]. It is a leading cause of sudden cardiac death (SCD) in young individuals and a common cause of heart failure in adults [3]. A hallmark feature specific to patients who die suddenly with HCM is extensive fiber disarray, which is thought to play a key role in the development of re-entrant ventricular arrhythmias leading to SCD [4]. This disorganization of myocardial fibers can be localized to specific regions of the left ventricle (LV) or spread throughout the entire LV [5]. In addition to fiber disarray, HCM is also associated with other pathological features, including LV hypertrophy, altered myocardial contractility, and cardiac fibrosis. The emergence of fiber disarray is often thought to be connected to contractile imbalance and heterogeneous myocardial stiffening due to fibrosis; however, the underlying mechanisms are still not fully understood.

Several studies have proposed that contractile imbalance, resulting from heterogeneity in force generation among cardiomyocytes, may contribute to the development of fiber disarray in patients with HCM [6-11]. To date, over 1,000 causative mutations have been identified for HCM, predominantly within genes encoding sarcomeric proteins responsible for generating and regulating contraction, with approximately one-third of these mutations located in β-cardiac myosin [12]. Experimental studies have shown substantial differences in force generation between individual cardiomyocytes from HCM patients with mutations in β-myosin heavy chain (β-MyHC) even at the same calcium concentration [7, 9, 10]. It is assumed that unequal fractions of mutant and wild-type protein, from cell to cell, may underlie such contractile imbalance. While the prevailing trend is that HCM mutations cause hypercontractility—linked to mutation-induced disruption of the super-relaxed state (SRX) of myosin heads [7, 13-16]—several studies have also reported reduced force generation in certain HCM mutations, leading to hypocontractility [7, 14, 17-19].

Alongside contractile imbalance, myocardial fibrosis plays a critical role in HCM pathology. Increased myocardial fibrosis is maladaptive, as its accumulation is correlated with impaired cardiac relaxation and a higher risk of heart failure [20]. Myocardial fibrosis can be categorized into the general subtypes of replacement fibrosis and interstitial fibrosis. In HCM, the predominant type of myocardial fibrosis is replacement fibrosis, and late gadolinium enhancement (LGE) with cardiac MRI is the reference standard for detecting this type of fibrosis [21-23]. The emergence of replacement fibrosis in HCM is driven by the premature death of mutant myocytes and the expansion of the interstitial matrix. Factors such as compromised coronary flow due to hypertrophy, microvascular dysfunction, increased oxidative stress, and heightened metabolic demands from abnormal biophysical properties of mutant sarcomeres contribute to myocyte death and the formation of replacement fibrosis in HCM [20]. The expansion of the extracellular matrix in fibrotic areas causes local stiffening and makes the myocardium highly heterogeneous, disrupting its mechanics and altering the stress distribution, which is assumed to contribute to further remodeling of myocardial fibers.

The intricate mechanisms underlying cardiac pathophysiological conditions are difficult to evaluate in a clinical setting due to ethical and technical limitations [24]. But, computational methods, particularly finite element (FE) modeling, offer a promising solution by overcoming these limitations and providing valuable insights into cardiac structure and function. A key aspect of these models is the effective estimation of the cardiac fiber configuration. Many models rely on rule-based assignments of myofiber angles based on experimental data [25-29], while others use warping and projection methods to integrate geometry-specific myofiber angles from diffusion tensor magnetic resonance imaging (DTMRI) [30-32]. A machine learning based model, which was used to estimate myocardial stiffness, showed the importance of fiber orientation on capturing realistic global pumping function [33]. Given the crucial role of fiber organization in cardiac function and the disruptions caused by pathological conditions, it is essential to use a model that is flexible in assigning the initial fiber configuration and can predict gradual fiber remodeling in conditions like HCM.

Kroon et al. (2009) developed a cardiac FE model that adapts myofiber orientation to minimize shear strain, based on the assumption that shear strain disrupts the extracellular matrix (ECM) and leads to permanent fiber reorientation [34]. Similarly, Washio et al. (2015) used an adaptive model to predict myofiber orientation in healthy hearts, based on two independent optimization parameters: (1) local muscle workload and (2) contractile load impulse. Both of these parameters were functions of "active" cardiac contraction, which limited their ability to reflect passive behavior, especially in fibrotic regions [35]. Washio et al. (2020) later found that the impulse-based mechanism was more suitable for myofiber reorientation [36]. Another study utilized an agent-based model to describe fibroblast-driven collagen remodeling, regulated by chemical and mechanical cues. This cell-level model was then scaled up to a 3D FE model of the LV, but was limited to simulating only the diastolic phase [37]. Mendiola et al. [38] used a combination of image-driven micromechanics and computational modeling to assess the effects of fiber orientation and fibrosis over time in a rat model of myocardial infarction. Other studies have investigated stress-based laws for collagen remodeling in arterial tissue [39, 40].

The current study employs a stress-based fiber remodeling algorithm that extends the previous methods by incorporating both active and passive stress as fiber adaptation drivers. This allows for the simulation of remodeling throughout the entire cardiac cycle in both myofibers and the extracellular matrix. In this approach, fibers tend to reorient toward local traction vectors, which can be altered due to heterogenous mechanical properties. Unlike earlier HCM finite element modeling studies that examine fiber disarray as a model input affecting cardiac performance [5, 41, 42], i.e. the disarray is assigned a priori, the current research investigates the underlying mechanisms driving fiber reorientation and predicts subsequent myocardial disarray. In a prior study, we demonstrated the accuracy of a stress-based fiber reorientation law in predicting both normal fiber configurations and pathological remodeling post myocardial infarction (MI) [43]. The aim of the current study is to deepen our understanding of HCM progression by examining how heterogeneous hypercontractility, hypocontractility, and fibrosis influence myocardial remodeling and cardiac function.  This is motivated by studies looking at the established stages of HCM, where the disease may be self-sustaining beyond the effects of the mutant protein affecting contractility, with the diseased ECM in fibrotic tissue potentially playing a significant role [44]. By isolating the effects of these cell-level HCM-induced abnormalities, this study provides new insights into disease progression and may help the development of more targeted therapeutic strategies.

# 2. Methods

## 2.1 MyoFE framework

The current study examines fiber remodeling and cardiac behavior related to HCM using a modular multiscale FE framework known as MyoFE (Figure 1). The MyoFE framework combines various aspects of cardiovascular function, including calcium dynamics, myofiber contraction, and circulatory hemodynamics, within a closed-loop model of the LV and systemic circulation. Additionally, a fiber reorientation module is coupled with this central framework to predict the fiber configuration in both the normal LV and adverse remodeling caused by HCM. All model parameters are detailed in Table S1 of the Supplementary Material.

Myofiber contraction is implemented using a molecular-level model called MyoSim [12, 26], which computes the active stress generated by sarcomeres based on the deformation of the myofibers and the calcium concentration. In this study, a pacing stimulus activates a two-compartment model of calcium dynamics. The mechanical configuration of the myofibers, including changes in sarcomere length and total stress, is derived from the 3D FE model of the LV. To impose the hemodynamic boundary condition related with the volume of the LV cavity, a Windkessel circulation model [45] is used. Essentially, the total stress state in the myocardium of the 3D FE model (i.e., passive and active stress) is balanced by the pressure in the LV cavity. The coupling of the FE model and circulatory model constrains the LV cavity volume calculated by each model, via a Lagrange multiplier, which allows for the determination of the cavity pressure.

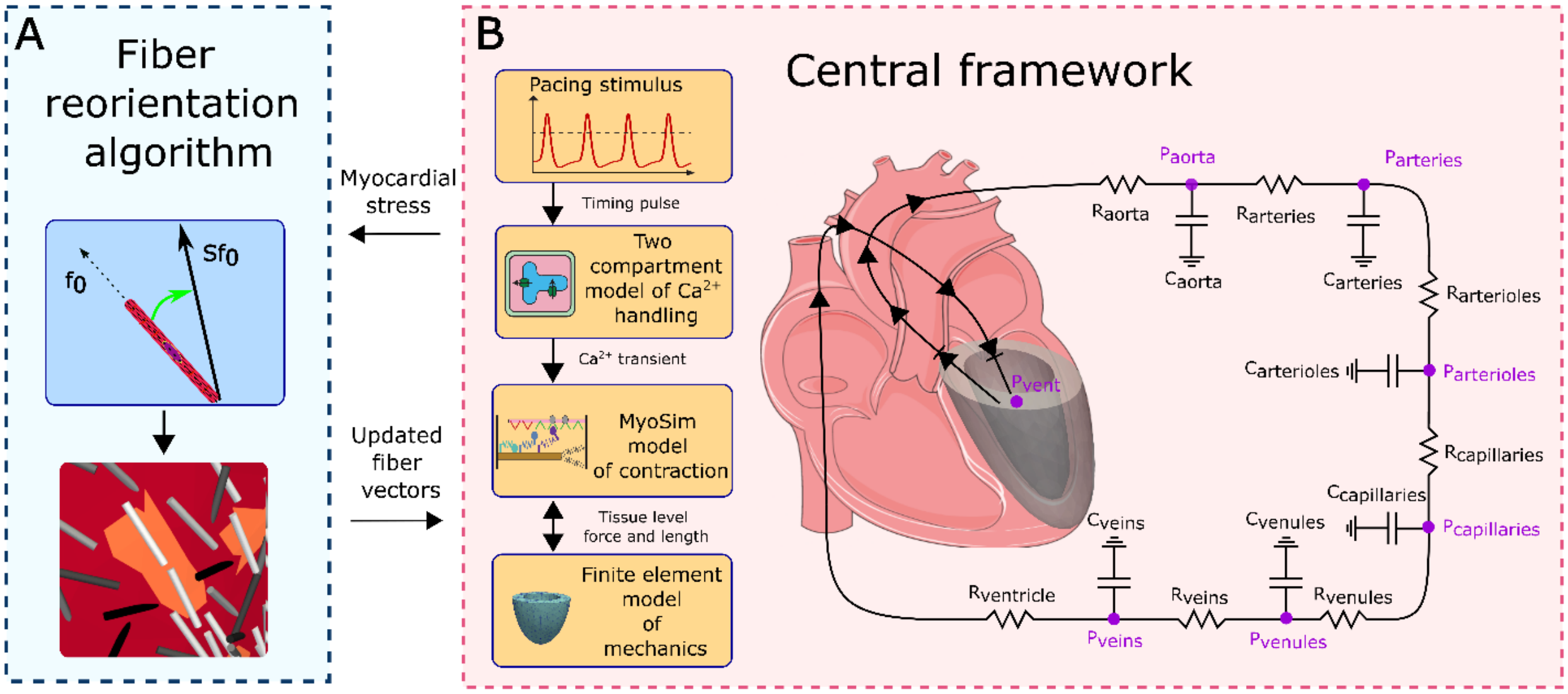

**Figure 1:** MyoFE framework schematic. **A:** The fiber reorientation algorithm reorients fibers using the myocardial stress from the LV model (upper panel) in the central framework. In regions of heterogeneous mechanical properties, this can lead to localized remodeling of the fibers (lower panel). The updated fiber vectors are then passed to the central framework for the subsequent time step. The unit vector $\boldsymbol{f_0}$ represents the initial direction of a fiber. $\boldsymbol{f_0}$ reorients toward the local traction vector $\boldsymbol{Sf_0}$, where $\boldsymbol{S}$ is the total stress tensor, encompassing both passive and active stresses. **B:** The central framework consists of the lumped parameter circulatory model and modules in the yellow boxes on the left. Detailed information for each module is provided in the text.

## 2.2 Circulation model

The FE model is joined with a Windkessel model of the closed systemic circulation to determine the hemodynamic boundary conditions within the LV cavity. This circulation model, which includes six compartments in addition to the left ventricle, is represented by a set of resistance (R) and compliance (C) parameters. In each compartment, the rate of change in blood volume is determined by the difference between the inflow and outflow of blood per unit time, as follows:

$$\begin{aligned}
\frac{dV_{aorta}}{dt} &= Q_{LV\ to\ aorta} - Q_{aorta\ to\ arteries} \\
\frac{dV_{arteries}}{dt} &= Q_{aorta\ to\ arteries} - Q_{arteries\ to\ arterioles} \\
\frac{dV_{arterioles}}{dt} &= Q_{arteries\ to\ arterioles} - Q_{arterioles\ to\ capillaries} \\
\frac{dV_{capillaries}}{dt} &= Q_{arterioles\ to\ capillaries} - Q_{capillaries\ to\ venules}\ . \\
\frac{dV_{venules}}{dt} &= Q_{capillaries\ to\ venules} - Q_{venules\ to\ veins} \\
\frac{dV_{veins}}{dt} &= Q_{venules\ to\ veins} - Q_{veins\ to\ LV} \\
\frac{dV_{LV}}{dt} &= Q_{veins\ to\ LV} - Q_{LV\ to\ aorta}
\end{aligned} \tag{1}$$

According to Ohm's law, the blood flow from one compartment to another is equal to the pressure gradient between adjacent compartments divided by the resistance between them:

$$Q_{LV\ to\ aorta} = \begin{cases} \frac{P_{LV}-P_{aorta}}{R_{aorta}} & when\ P_{LV} \geq P_{aorta} \\ 0 & otherwise \end{cases}$$
$$Q_{aorta\ to\ arteries} = \frac{P_{aorta}-P_{arteries}}{R_{arteries}}$$
$$Q_{arteries\ to\ arterioles} = \frac{P_{arteries}-P_{arterioles}}{R_{arterioles}}$$
$$Q_{arterioles\ to\ capillaries} = \frac{P_{arterioles}-P_{capillaries}}{R_{capillaries}} \quad . \quad (2)$$
$$Q_{capillaries\ to\ venules} = \frac{P_{capillaries}-P_{venules}}{R_{venules}}$$
$$Q_{venules\ to\ veins} = \frac{P_{venules}-P_{veins}}{R_{veins}}$$
$$Q_{veins\ to\ LV} = \begin{cases} \frac{P_{veins}-P_{LV}}{R_{LV}} & when\ P_{veins} \geq P_{LV} \\ 0 & otherwise \end{cases}$$

For all compartments except the LV, the blood pressure is determined by the difference between the instantaneous and slack volumes divided by the compartment's compliance:

$$P_i(t) = \frac{V_i(t)-V_{i,slack}}{C_i}\ , \quad (3)$$

where $V_i(t)$, $V_{i,slack}$ and $C_i$ are the instantaneous blood volume, the slack volume, and the compliance of compartment i, respectively. For the LV, cavity pressure is determined via the FE model, using the LV cavity volume from equation (1) as a boundary condition as described below.

### 2.3 Finite element formulation

The LV mechanics in this study are solved using a FE model implemented with the open-source FE library FEniCS [46]. An implicit backward Euler scheme with a fixed time step of 1 ms is used for time integration. The displacement field is represented by quadratic interpolation over the mesh, while linear interpolation is used to describe the hydrostatic pressure. The FE formulation governs the LV mechanics based on the minimization of the Lagrangian functional as described below:

$$\begin{aligned} \mathcal{L}(\boldsymbol{u}, p, P_{LV}, c_x, c_y, c_z) = \int_{\Omega_0} W(\boldsymbol{u})\, dV - \int_{\Omega_0} p(J-1)\, dV - \\ P_{LV}(V_{LV}(\boldsymbol{u}) - V_{LV}) - c_x \cdot \int_{\Omega_0} u_x\, dV - \\ c_y \cdot \int_{\Omega_0} u_y\, dV - c_z \cdot \int_{\Omega_0} \boldsymbol{z} \times \boldsymbol{u}\, dV \end{aligned} \quad , \quad (4)$$

where $\boldsymbol{u}$ is the displacement field, $W$ is the total strain energy, and $p$ is a Lagrange multiplier that enforces incompressibility by constraining the Jacobian of the deformation gradient tensor to unity (J=1). The LV cavity volume $V_{LV}(\boldsymbol{u})$ is constrained via the Lagrange multiplier $P_{LV}$ according to the hemodynamic boundary condition from the circulatory model $V_{LV}$. Lagrange multipliers $c_x$ and $c_y$ constrain rigid body translation in the **x** and **y** directions, respectively, and $c_z$ constrains rigid body rotation.

Additionally, the relationship between the volume of the LV cavity and the displacement field is defined as:

$$V_{LV}(\boldsymbol{u}) = \int_{\Omega_{k,endo}} dV = -\frac{1}{3}\int_{\Gamma_{k,endo}} \overline{\boldsymbol{x}} \cdot \boldsymbol{n}\, da \quad , \quad (5)$$

where $\Omega_{k,endo}$ denotes the volume enclosed by the endocardial surface $\Gamma_{k,endo}$ and the basal plane at z = 0 and $\boldsymbol{n}$ is the direction of the outward normal vector to the surface. The position vector $\overline{\boldsymbol{x}}$ is relative to the origin of the global coordinate system.

Taking the first variation of equation (4) allows us to express the weak formulation of the mechanics problem as follows:

$$\delta\mathcal{L}(\boldsymbol{u}, p, P_{LV}, c_x, c_y, c_z) = \int_{\Omega_0} \boldsymbol{FS}: \nabla\boldsymbol{\delta u}\, dV - \int_{\Omega_0} (pJ\boldsymbol{F}^{-\boldsymbol{T}}: \nabla\boldsymbol{\delta u} - \delta p(J-1)\, dV - P_{LV}\int_{\Omega_0} J\boldsymbol{F}^{-\boldsymbol{T}}: \nabla\boldsymbol{\delta u}\, dV - \delta P_{LV}(V_{LV}(\boldsymbol{u}) - V_{LV}) - \delta c_x \cdot \int_{\Omega_0} u_x\, dV - c_x \cdot \int_{\Omega_0} \delta u_x\, dV - , \quad (6)$$
$$\delta c_y \cdot \int_{\Omega_0} u_y\, dV - c_y \cdot \int_{\Omega_0} \delta u_y\, dV - \delta c_z \cdot \int_{\Omega_0} \boldsymbol{z} \times \boldsymbol{u}\, dV - c_z \cdot \int_{\Omega_0} \boldsymbol{z} \times \boldsymbol{\delta u}\, dV = 0$$

where $\boldsymbol{F}$ denotes the deformation gradient tensor, $\boldsymbol{S}$ is the 2nd Piola Kirchhoff stress tensor, $\boldsymbol{\delta u} \in H^1(\Omega_0)$, $\delta p \in L^2(\Omega_0)$, $\delta P_{LV} \in R$, $\delta c_x \in R$, $\delta c_y \in R$, and $\delta c_z \in R$ are test functions corresponding to $u$, $p$, $P_{LV}$, $c_x$, $c_y$, and $c_z$, respectively. In addition, a Dirichlet boundary condition $\boldsymbol{u} \cdot \boldsymbol{n}|_{base} = 0$ is considered for the base of the LV, which enforces in-plane deformation. The pressure in the LV cavity, derived from the Lagrange multiplier $P_{LV}$ is incorporated into circulatory equation (2), ensuring complete coupling between the 3D FE model and the circulatory model.

### 2.4 Mechanics of cardiac muscle

This study incorporates both the passive and active properties of the myocardium to model the mechanical behavior of the LV. Accordingly, the 2nd Piola Kirchhoff stress tensor is additively decomposed into active ($\boldsymbol{S_a}$) and passive ($\boldsymbol{S_p}$) components:

$$\boldsymbol{S} = \boldsymbol{S_a} + \boldsymbol{S_p} \, . \quad (7)$$

To capture the force-dependent nature of myofibers, it is essential to decompose the passive response into two distinct components: the myofiber component and the remaining bulk material. Furthermore, a passive component is incorporated to account for the incompressibility of the tissue. The passive stress tensor for each component is calculated by taking the derivative of the strain energy function with respect to the Green-Lagrange strain tensor $\boldsymbol{E}$:

$$\boldsymbol{S_p} = \boldsymbol{S_{vol}} + \boldsymbol{S_{bulk}} + \boldsymbol{S_{myofiber}} = \frac{\partial\Psi_{vol}}{\partial\boldsymbol{E}} + \frac{\partial\Psi_{bulk}}{\partial\boldsymbol{E}} + \frac{\partial\Psi_{myofiber}}{\partial\boldsymbol{E}} \quad , \quad (8)$$

where the volumetric strain energy function is $\Psi_{vol} = -p(J-1)$, with a Lagrange multiplier $p$ to enforce incompressibility exactly. The strain energy function of the bulk tissue, which reflects the distribution of collagen, elastin, and microvasculature, is modeled as a transversely isotropic material using the Guccione constitutive law [47]:

$$\begin{gathered}\Psi_{bulk} = \frac{C}{2}(e^{Q}-1) \\ with\ Q = b_{ff}E_{ff}^2 + b_{xx}(E_{ss}^2 + E_{nn}^2 + E_{sn}^2 + E_{ns}^2) + b_{fs}\left(E_{fs}^2 + E_{sf}^2 + E_{fn}^2 + E_{nf}^2\right)\end{gathered}, \quad (9)$$

where $C$, $b_{ff}$, $b_{xx}$, and $b_{fs}$ are passive material parameters for the bulk tissue. ***E*** represents the Green-Lagrange strain tensor defined with respect to the fiber, sheet, and sheer-normal directions (*f, s,* and *n* components), respectively. Lastly, the myofiber strain energy function is defined as:

$$\Psi_{myofiber} = \begin{cases} C_1\left(e^{C_2(\alpha-1)^2}\right) & \alpha > 1 \\ 0 & \alpha \leq 1 \end{cases}, \quad (10)$$

where $\alpha = \sqrt{\boldsymbol{f_0} \cdot \boldsymbol{C}\boldsymbol{f_0}}$ is the stretch along the fiber direction, $\boldsymbol{C}=\boldsymbol{F}^T\boldsymbol{F}$ is the right Cauchy-Green deformation tensor, $\boldsymbol{f_0}$ is the referential fiber direction, and $C_1$, $C_2$ are material constants.

The active stress component ($\boldsymbol{S_a}$) is calculated using the MyoSim framework [12], which models the mechanical properties of dynamically coupled myofilaments of a half sarcomere at each integration point in the LV FE mesh. A schematic of the cross-bridge scheme is illustrated in Figure 2 [28]. To describe the MyoSim methodology, we define the transition fluxes of the binding sites on the actin filaments, $J_{on}$ and $J_{off}$ and then describe the transition fluxes for the myosin filaments, represented by $J_1$, $J_2$, $J_3$, and $J_4$, which capture various states of the myosin heads [26].

The binding sites on actin transition between an inactive state ($N_{off}$), where no myosin heads can attach, and an active state ($N_{on}$). The binding sites that are active can be categorized into two configurations: $N_{unbound}$, representing sites that are not bound to myosin heads, and $N_{bound}$, representing sites that are bound to myosin heads and cannot transition back to the $N_{off}$ state. $J_{on}$ is the flux that governs the transition of binding sites from $N_{off}$ to $N_{on}$ and is defined as:

$$J_{on} = k_{on}[Ca^{2+}](N_{overlap} - N_{on})\left(1 + k_{coop}\left(\frac{N_{on}}{N_{overlap}}\right)\right), \quad (11)$$

where $N_{on}$ represents the proportion of binding sites in the active state, $k_{on}$ denotes a rate constant, $N_{overlap}$ represents the proportion of binding sites in the vicinity of myosin heads, $k_{coop}$ is a constant factor that controls the cooperativity of the thin filament.

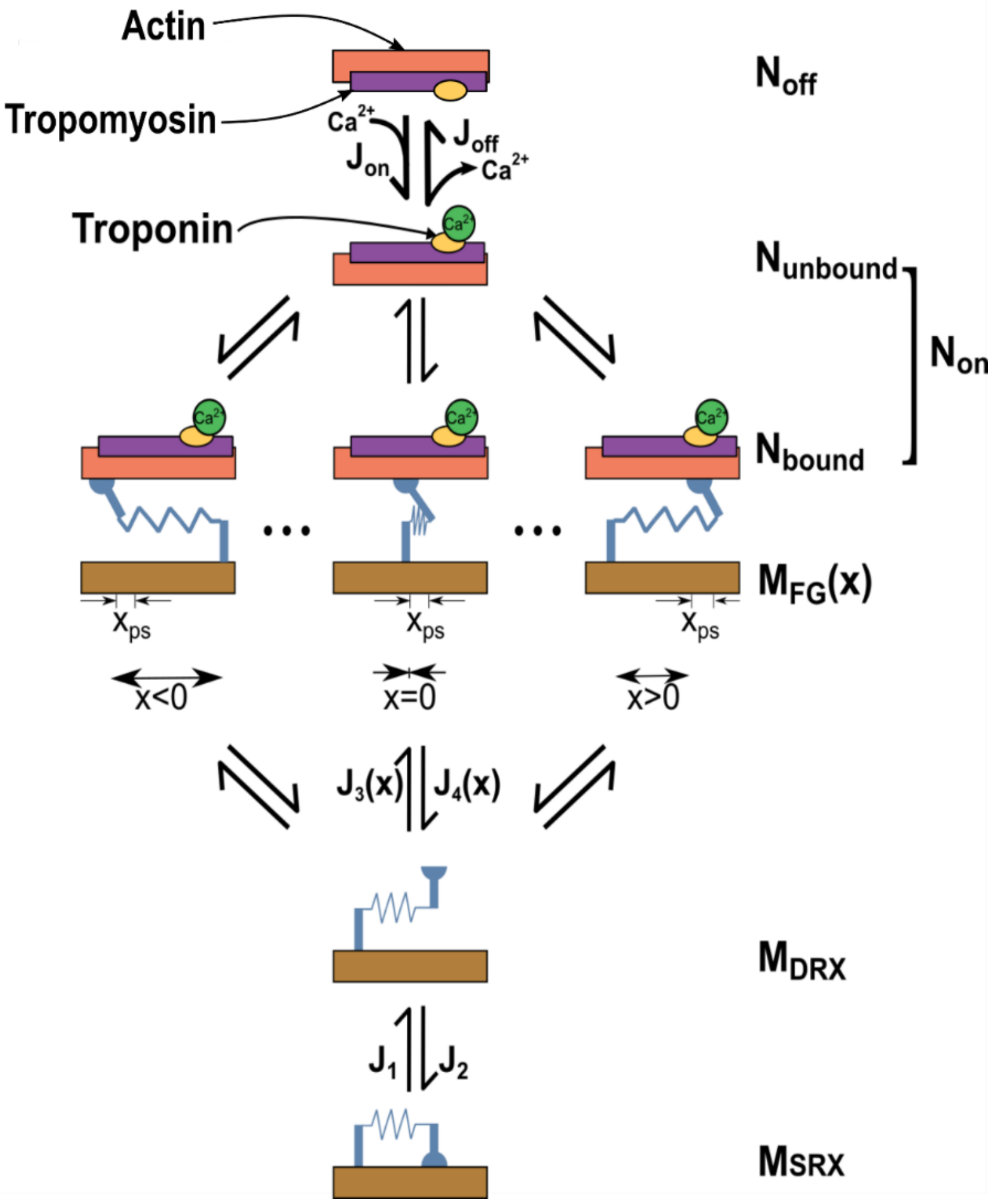

**Figure 2:** Cross-bridge scheme. Sites on the thin filament switch between states that are available ($N_{on}$) and unavailable ($N_{off}$) for cross-bridges to bind to. Myosin heads transition between a super-relaxed detached state ($M_{SRX}$), a disordered-relaxed detached state ($M_{DRX}$), and a single attached force-generating state ($M_{FG}$). J terms indicate fluxes between different states. Reproduced with permission from Springer Nature from Sharifi et al [28].

The activated unbound sites transition back into the inactive state via the $J_{off}$ flux, which is governed by a rate constant of $k_{off}$:

$$J_{off} = k_{off}\,(N_{on} - N_{bound})\left(1 + k_{coop}\left(\frac{N_{overlap} - N_{on}}{N_{overlap}}\right)\right) \quad . \tag{12}$$

On the myosin filament, myosin heads can switch between three states: $M_{SRX}$ (detached super-relaxed), $M_{DRX}$ (detached disordered-relaxed), and $M_{FG}$ (attached force generating). The fluxes governing the transitions between $M_{SRX}$ and $M_{DRX}$ are represented by $J_1$ and $J_2$, respectively as below:

$$\begin{aligned} J_1 &= k_1\,(1 + k_{force}F_{total})M_{SRX} \quad . \\ J_2 &= k_2 M_{DRX} \end{aligned} \tag{13}$$

In this equation, $k_1$ and $k_2$ are the rate constants, $k_{force}$ is a parameter related to force-dependent recruitment of myosin heads, $F_{total}$ is the total stress (passive and active) along the myofiber, and $M_{SRX}$ and $M_{DRX}$ are the proportions of myosin heads in the SRX and DRX states. Notably, since the HCM mutations disrupt the SRX state, we implemented the abnormal contractility of the diseased LV by adjusting $k_1$.

The attachment of myosin heads to binding sites is defined by the transition flux $J_3$, while the detachment is governed by the transition flux $J_4$:

$$J_3\,(x) = k_3 e^{\frac{-k_{cb}x^2}{2k_B T}} (N_{on} - N_{bound}) M_{DRX} \quad , \tag{14}$$

$$J_4\,(x) = (k_{4,0} + k_{4,1}x^4)\, M_{FG}(x)$$

where $k_3$ and $k_{4,0}$ are rate constants, $k_{cb}$ is the stiffness of the cross-bridge link, $k_B$ is the Boltzmann constant, $T$ is temperature in Kelvin, $k_{4,1}$ is the strain dependent cross-bridge detachment rate parameter, $M_{FG}$ is the proportion of myosin heads in the attached state, and $x$ is the cross-bridge length.

By utilizing the transition fluxes mentioned above, a system of ordinary differential equations (ODEs) is created to determine the proportion of myosin heads and binding sites in each configuration. This system of ODEs is partitioned with a spatial resolution of 1 nm, spanning a range of $-10 \leq x \leq 10$ nm (i.e., n=21). As a result, a total of 25 ODEs are solved at each integration point within the LV mesh, defined as:

$$\begin{aligned} \frac{dN_{off}}{dt} &= -J_{on} + J_{off} \\ \frac{dN_{on}}{dt} &= J_{on} - J_{off} \\ \frac{dM_{SRX}}{dt} &= -J_1 + J_2 \\ \frac{dM_{DRX}}{dt} &= \left(J_1 + \sum_{i=1}^{n} J_{4,x_i}\right) - \left(J_2 + \sum_{i=1}^{n} J_{3,x_i}\right) \\ \frac{dM_{FG,i}}{dt} &= J_{3,x_i} - J_{4,x_i}\ where\ i = 1 \ldots n \end{aligned} \quad . \tag{15}$$

The active stress experienced along each myofiber can be determined by using the proportion of myosin heads in the force-generating state, expressed as:

$$F_{active} = N_o k_{cb} \sum_{i=1}^{n} M_{FG,i}\,\left(x_i + x_{ps}\right) \quad , \tag{16}$$

where $N_0$ represents the density of myosin heads [48], and $x_{ps}$ refers to the power stroke of an attached cross-bridge.

Finally, the active stress tensor is defined by projecting $F_{active}$ along the myofiber direction:

$$\boldsymbol{S_a} = F_{active}\boldsymbol{f_0} \otimes \boldsymbol{f_0} \quad . \tag{17}$$

## 2.5 Stress-Based Fiber Reorientation

The fiber reorientation mechanism is depicted in Figure 3, where $\boldsymbol{f_0}$ represents the unit vector along the fiber direction in the reference configuration, which is shared by the myofibers and perimysial collagen fibers. The traction vector ($\boldsymbol{Sf_0}$) is associated with the cross-sectional face of the fibers at a point in the LV. The stress-based reorientation law incrementally reorients the fibers at each integration point in the LV mesh, aligning them with the local traction vector as below:

$$\frac{d\boldsymbol{f_0}}{dt} = \frac{1}{\kappa}\left(\frac{\boldsymbol{Sf_0}}{\|\boldsymbol{Sf_0}\|} - \boldsymbol{f_0}\right) \quad , \qquad (22)$$

where $\kappa$ is a time constant used to enforce the separation of time scales, which acts like a map between model time and real time. This is a numerical necessity to speed up the calculation of remodeling. Specifically, it facilitates a reasonable rate of fiber reorientation, enabling the capture of case-specific fiber configurations within a limited number of cardiac beats, which leads to better computationally efficiency. The gradual reorientation process minimizes the difference between the fiber direction and the direction of the traction experienced by the fibers. To capture the reorientation occurring in both myofibers and collagen, it is necessary to use the total stress tensor, which includes both active and passive stress components. This is especially important in the fibrotic regions where the scar tissue is devoid of contracting myofibers. For more details on the stress-based fiber reorientation law, please refer to our previous paper [43].

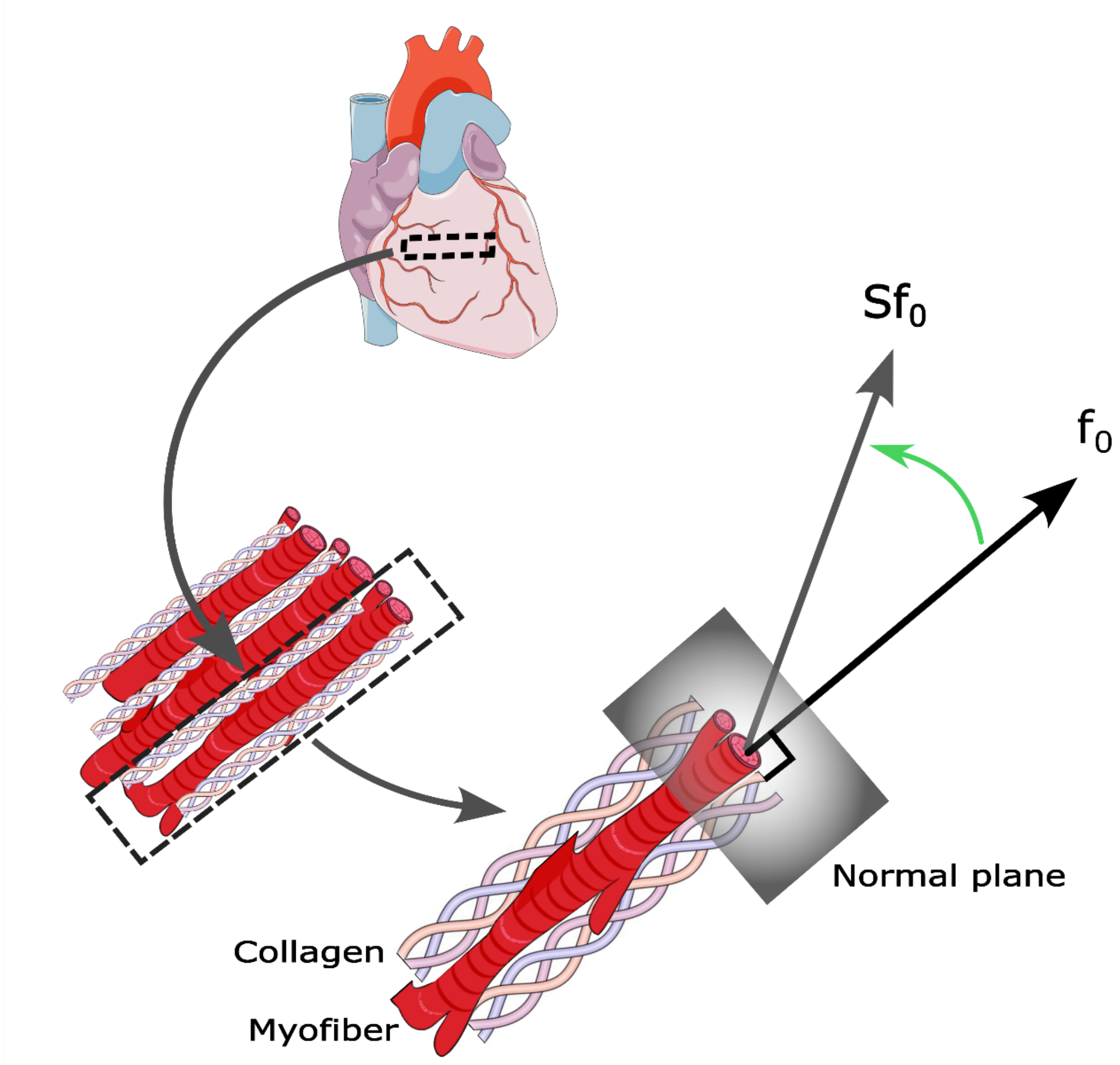

**Figure 3:** Stress based fiber reorientation. The unit vector $\boldsymbol{f}_0$ represents the initial direction of the myofiber and collagen. $\boldsymbol{f}_0$ reorients toward the local traction vector $\boldsymbol{S}\boldsymbol{f}_0$, which is associated with the cross-sectional face of the fiber. $\boldsymbol{S}$ is the total stress tensor, encompassing both passive and active stresses. Reproduced with permission from Elsevier from Mehri et al [43].

### 2.6 LV model geometry

The LV of a human heart was modeled using an ellipsoidal geometry consisting of ~1250 quadratic tetrahedral elements (refer to Supplementary Figure S1 for the mesh sensitivity analysis). The LV's unloaded chamber volume was 75 mL, with a length from base to apex of 7.3 cm. At the base, the outer diameter was 7.2 cm, and the wall thickness was about 1.3 cm, while the apical wall thickness was half of that value [25]. The initial helical fiber angle was defined using a Laplace-Dirichlet rule-based algorithm, varying linearly from 60° at the endocardium to -60° at the epicardium transmurally across the wall's thickness [49]. The initial transverse angle was set to zero.

### 2.7 Simulation cases and protocol

To investigate the distinct effects of cell-level abnormalities induced by HCM, we developed a baseline LV model, as well as hypercontractile, hypocontractile, and fibrotic models. Previous studies have shown that mutant myocytes undergo biophysical damage, leading to myocyte death and focal scarring (replacement fibrosis) [20]. Therefore, we hypothesized that the myocardial regions exhibiting abnormal contractility are the same regions that progress to fibrosis and assume that these regions are similarly distributed in all perturbed models. We utilized the results from histological studies of fibrotic tissue to prepare our fibrous model and applied the same distribution pattern to the hypercontractile and hypocontractile models, as shown in Figure 4A. This assumption enables a direct comparison of how each type of abnormality affects the simulation results. Our assumption regarding the distribution of perturbed regions in the models is also inspired by T2* images, which show patterns of regional abnormal contractility in HCM patients. Myocardial T2* is commonly regarded as a surrogate for myocardial tissue oxygenation and is associated with myocardial contraction [50].

According to measurements by Galati et al., we incorporated heterogeneous perturbed tissue consisting of 30% of the LV myocardium in all of the diseased LV models [21]. Notably, based on histological studies, fibrosis in HCM is almost evenly distributed among the epicardium, endocardium, and midwall layer [21]. Therefore, we did not assign any regional preferences, and the perturbed cells were randomly distributed within the myocardium of the LV.

To determine whether variations in the distribution and size of perturbed regions influence the severity and pattern of fiber disarray, we modeled all cases using two additional different distributions of perturbed cells. According to prior histological studies, collagen scars of uninterrupted fibrosis, characterized by local foci greater than 2 mm, are classified as replacement fibrosis [21]. Accordingly, we investigated three different sets of models with perturbed regions of minimum sizes: 3 mm for Distribution 1 (original distribution), 2.5 mm for Distribution 2, and 2 mm for Distribution 3 as shown in Figure 4 B.

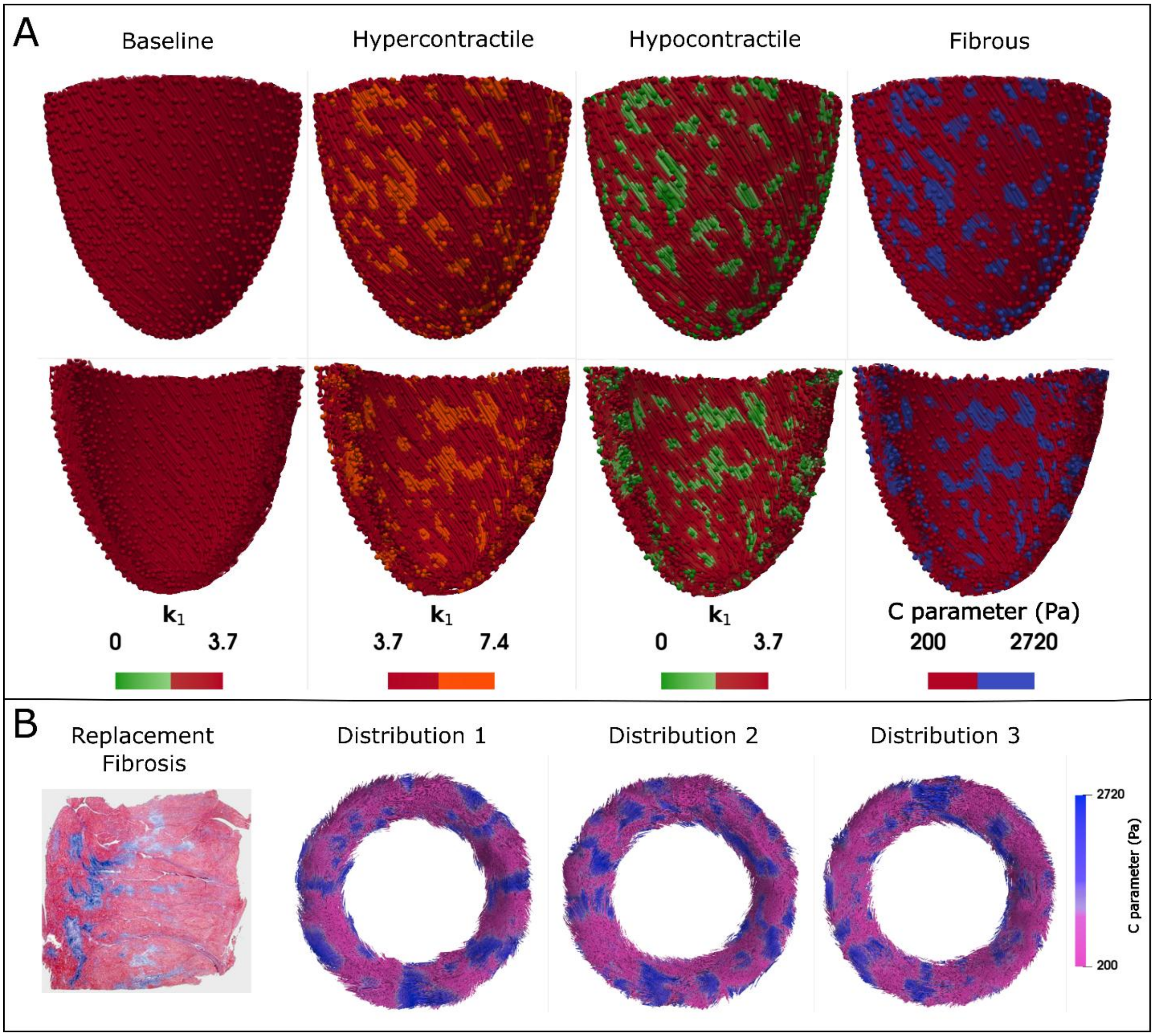

**Figure 4:** LV models. **A:** Epicardial and endocardial views of the baseline and perturbed LV models. For the baseline and hyper/hypocontractile models, the $k_1$ parameter represents myocardial contractility, while in the fibrous model, the C parameter represents stiffness. The spheres represent the integration points where properties are defined, while the color map represents the interpolated heterogenous regions. **B:** From left to right, a histological cross-section of myocardium with replacement fibrosis excised from a patient with HCM (adapted from [51]), followed by short-axis views of LV models at mid-ventricle with varying distributions of perturbed regions. Distribution 1 represents perturbed regions of minimum size 3 mm, followed by Distribution 2 (2.5 mm) and Distribution 3 (2 mm). All result figures in this paper are based on Distribution 1, and other distributions are investigated in Section 3.7.

### Baseline

A baseline simulation was developed using parameters that are representative of a healthy adult, with cardiac performance characteristics aligned with values reported in the literature [52, 53]. The total blood volume was set to 4.5 liters, the end-diastolic volume (EDV) was 123 ml, the end-systolic volume (ESV) was 54 ml, and the ejection fraction (EF) of the LV was approximately 56% with a heart rate of 66 beats per minute. Alongside the baseline LV model representing a healthy case, we prepared perturbed LV models as described below:

**Hypercontractile**

Considering the disruption of the SRX state of myosin heads in HCM patients, which modulates myocardial contraction [7, 12], we incorporated a 100% increase in the $k_1$ value for perturbed cells in the hypercontractile models, where $k_1$ represents the transition rate of SRX to DRX in the MyoSim contraction model. Subsequently, we investigated the sensitivity to $k_1$ values by testing models with increased values of 20%, 60%, and 100%.

**Hypocontractile**

For the hypocontractile models, $k_1$ is reduced with a similar percentage to the hypercontractile models in the affected cardiomyocytes with impaired contractility.

**Fibrous**

To model the effects of heterogeneous replacement fibrosis in the LV, perturbed myocardial regions are assigned with a 2-fold increase in stiffness, as well as being isotropic [44]. We utilized the stress-strain relation for biaxial loading within the physiological strain range (~15%) to identify the corresponding material parameters reflecting the increase in stiffness (see Supplementary Material Figure S2). The material properties of the fibrotic regions, incorporating both passive and active mechanical responses, were applied based on the parameters listed in Table 1.

***Table 1:*** *Regional material properties of cardiac muscle in the LV models for normal and fibrous myocardium.*

| | ***Passive properties*** | | | | | | ***Active properties*** |
|---|---|---|---|---|---|---|---|
| *Region* | C (Pa) | $C_1$ (Pa) | $C_2$ | $b_f$ | $b_t$ | $b_{fs}$ | Cross-bridge density ($m^{-2}$) |
| *normal* | 200 | 250 | 15 | 8.0 | 3.58 | 1.63 | 6.96e+16 |
| *fibrous* | 2720 | 0 | 0 | 4 | 4 | 4 | 0 |

**Protocol**

The adaptation time constant, $\kappa$, is set to 4000 ms (Supplementary Material). The cardiac cycle time is assigned to be 937 ms (mimicking 64 bpm), with a numerical integration time step of 1 ms. Therefore, during each simulation time step, a 2.5x10$^{-4}$ (dt/$\kappa$) fraction of the difference between the traction vector and fiber vector is used to update the fiber direction. All models ran for 200 seconds (211 cardiac cycles) to ensure convergence (cardiac function reached a steady state, as shown in the pressure-volume (PV) loops in Supplementary Material Figure S3) and results are reported for the last cardiac cycle. The simulation protocol for all the LV models were developed with the following steps: (1) the initial fiber configuration was assigned to the LV using a rule-based approach, as described in Section 2.6, then (2) the system reached hemodynamic steady state by simulating the first five seconds without permitting fiber reorientation, finally (3) fiber reorientation was simulated over the remaining simulation time (195 s) to obtain the case-specific fiber configuration.

### 2.8 Implementation and Computer Code

The MyoFE source code was developed in Python utilizing widely used libraries such as Numpy [54], Scipy [55], and pandas [56]. All simulations in this research were run on the Lipscomb Compute Cluster (LCC), a high-performance computing system at the University of Kentucky. The required time for simulating one cardiac cycle using a single node containing 32 cores was about 15 minutes. In this study, model parameters were not calibrated with specific experimental data; instead, the default parameters listed in Supplementary Material Table S1 were chosen based on prior studies and experience, ensuring the model replicates typical healthy human characteristics [45].

### 2.9 Code Availability

The MyoFE code is maintained in a github repository. The code will be made available to other researchers upon request to the corresponding author.

## 3. Results

### 3.1. Comparison of Stress State

To explore the behavior of all models leading to fiber remodeling, we began by examining stress distributions, which serve as the driver of our fiber reorientation algorithm. As shown in Figure 5, the baseline model exhibits smooth stress distributions across the endocardial surface and epicardial surface, with a decreasing transmural gradient from endo to epi, as well as variations from base to apex. This trend is seen in the LV during both peak systole and diastole. In contrast, all perturbed models displayed highly heterogeneous stress distributions. In Figure 5A, the hypercontractile model shows regions of elevated systolic stress, relative to baseline, while the hypocontractile and fibrous models exhibit a mix of elevated and suppressed active stress relative to baseline. Figure 5B reveals a different pattern for peak diastolic stress: the hypocontractile model demonstrates increased passive stress on the endocardial side, relative to baseline, which is most likely induced by an increase in end-diastolic volume (to be discussed in a later section). However, no such elevation is seen in the hypercontractile model and the fibrous model shows a reduction in stress. A similar, though milder, pattern is observed on the epicardial side.

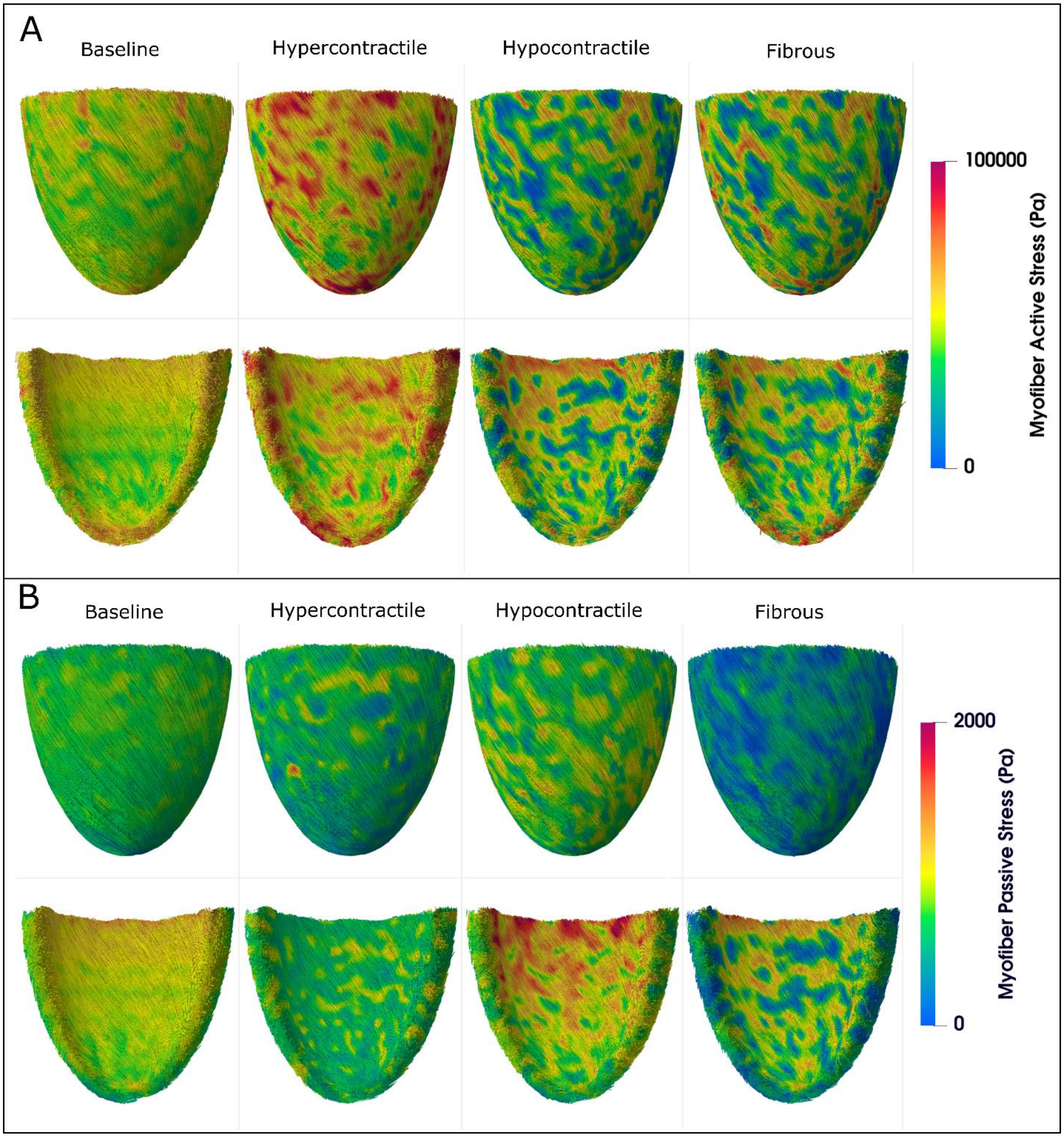

**Figure 5:** A comparison of myofiber stress distributions in the baseline and perturbed LV models. All models show the 2nd Piola Kirchhoff stress mapped onto the reference configuration. **A:** Myofiber active stress in peak systole (when ventricular volume is minimum just before relaxation). **B:** Myofiber passive stress in peak diastole (when ventricular volume is maximum just before contraction). Top row of each panel: Epicardial side of the myocardium. Bottom row of each panel: Endocardial side of the myocardium.

### 3.2. Comparison of fiber reorientation

This section evaluates the severity of fiber reorientation in each model and identifies the patterns of fiber disarray induced by each underlying perturbation. Figure 6 shows the magnitude of the 3D angle between the initial and reoriented fibers using the dot product of their direction vectors. In the baseline model, this value remains low and uniform, whereas in all perturbed models, fiber reorientation is significantly heterogeneous, indicating substantial fiber disarray.

On the epicardial surface (Figure 6A), the highest degree of fiber reorientation is observed in the fibrous model, slightly exceeding that of the hyper- and hypocontractile models. The magnified view of Figure 6A further emphasizes the regional severity and unique patterns of fiber disarray induced by each perturbation. In the hypercontractile model, fibers within the perturbed region show minimal reorientation, whereas the adjacent normal myocardium experiences severe reorientation. In contrast, both hypocontractile and fibrous models exhibit pronounced fiber reorientation within the perturbed region, accompanied by milder reorientation in the surrounding normal myocardium. This level of disarray, with fibers nearly orthogonal to each other, is consistent with previously reported findings in the literature for HCM (Figure 6B). On the endocardial surface (Figure 6.C), comparable patterns of fiber disarray are observed, with more pronounced fiber reorientation in the hypocontractile and fibrous models compared to the hypercontractile model. These results collectively demonstrate the distinct spatial patterns and severity of fiber disarray associated with each perturbation type, providing mechanistic insights into the development of myocardial fiber disarray in HCM. Additional visual representations of the fiber disarray for each case can be found in Supplementary Figure S4.

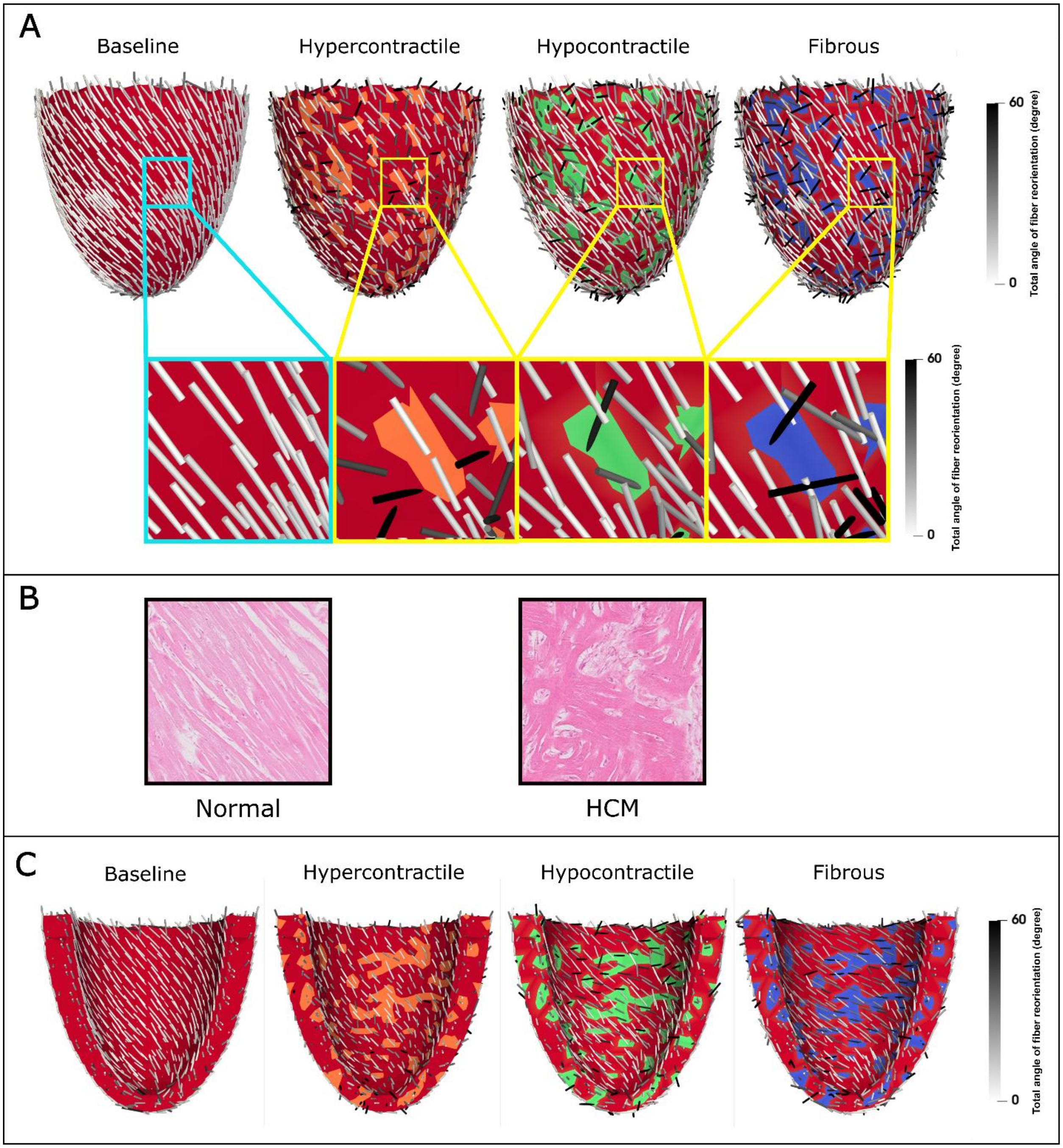

**Figure 6:** Patterns of fiber disarray caused by cell-level perturbations in HCM. (A) Total angle of fiber reorientation (magnitude of the 3D angle between initial and reoriented fibers) in the baseline and perturbed LV models, shown on the epicardial side of the myocardium. Magnified views provide detailed insights into regional fiber reorientation within and in the vicinity of the perturbed region. (B) Myocardial histology comparing normal and HCM cases (adapted with permission from Springer Nature from Matos al. [57]), illustrating fiber disarray in HCM as a loss of myocyte architectural organization, with fibers oriented obliquely or perpendicularly in a disorganized pattern [57]. (C) Total angle of fiber reorientation in the baseline and perturbed LV models, displayed on the endocardial side of the myocardium. In panels (A) and (C), red represents the normal myocardium, while the other colors represent the heterogeneous hypercontractile (orange), hypocontractile (green), and fibrous (blue) regions.

### 3.3. Comparison of Fiber Disarray

As a quantitative measure of fiber disarray, the angular deviation (AD) of the helical and transverse angles were analyzed as shown in Figure 7. AD is defined as the regional standard deviation within the epicardial layer and endocardial layer, each with a thickness of 20% of the transmural thickness. The baseline model exhibits relatively low AD, indicating minor variations in fiber orientations within the epicardial and endocardial layers. In all perturbed cases, helical AD (Figure 7A) shows a substantial increase in the epicardium, with a comparatively smaller rise in the endocardium. In the endocardium, the hypocontractile model exhibited the highest helical AD, followed by the hypercontractile and fibrous models. Similarly, transverse AD also increased across all perturbed models (Figure 7B), with higher values observed in the epicardium compared to the endocardium. Notably, within the endocardium, the hypercontractile model displayed the highest transverse AD, while the fibrous model showed the greatest value in the epicardial layer.

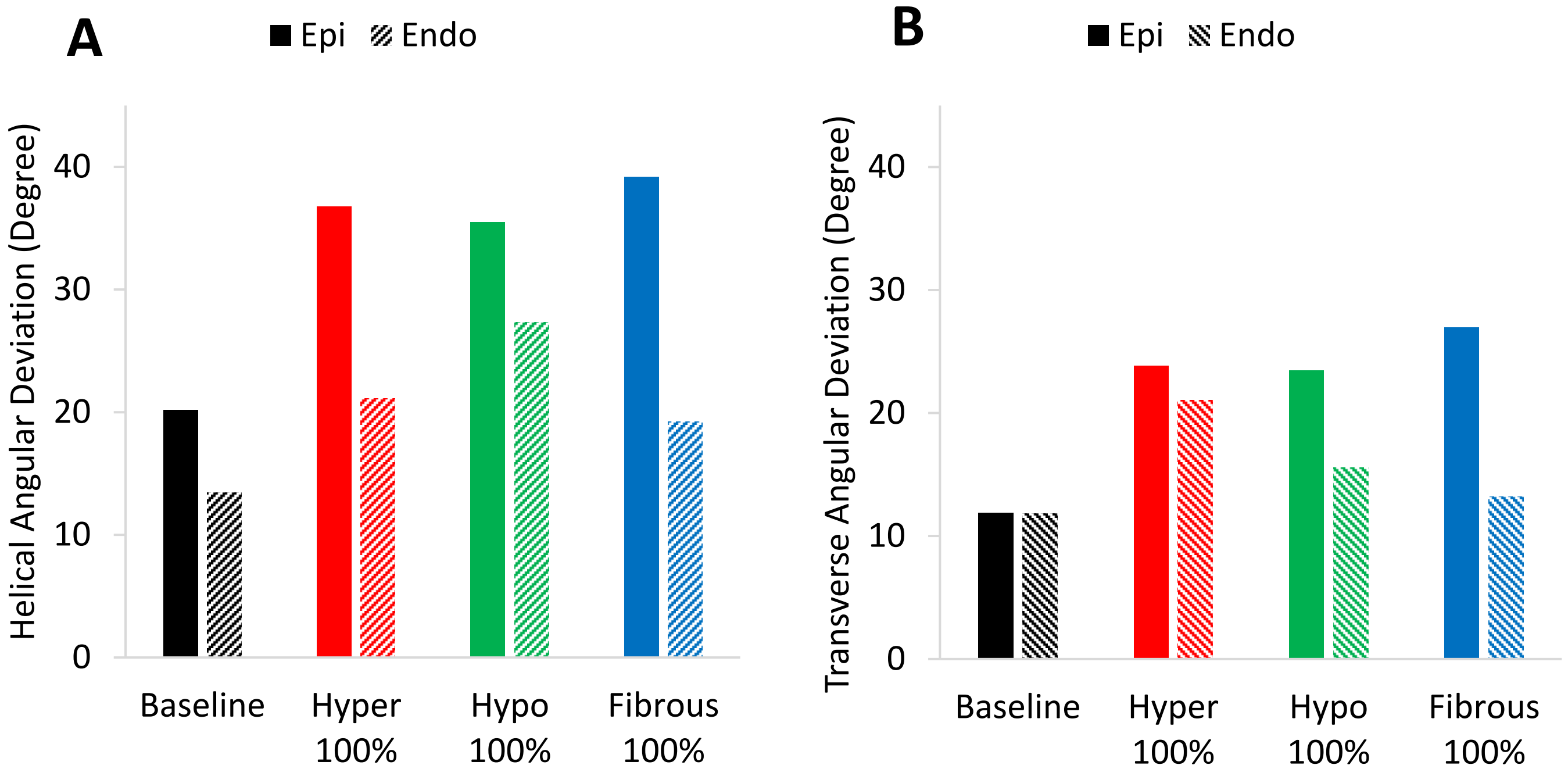

**Figure 7:** Angular deviations in the epicardium and endocardium across baseline and perturbed LV models **A:** Helical angular deviation. **B:** Transverse angular deviation

### 3.4. Contractile Sensitivity of Disarray

In this section, we investigate how the severity of abnormal contractility influences the degree of fiber disarray. For hypercontractility, we developed separate models with $k_1$ increase of 20%, 60%, and 100% from the baseline LV model value ($k_1 = 3.7$). Similarly, we created hypocontractile models with corresponding $k_1$ reduced by the same percentages. Notably, we assigned the distributions of perturbed cells to be the same across all models.

In all cases, we observed an almost linear trend between the induced fiber disarray (AD) and the $k_1$ parameter, as shown in Figure 8. Notably, all models demonstrated that fiber disarray is more pronounced near the epicardium compared to the endocardium. Near the epicardium, the hypercontractile and hypocontractile models exhibited similar sensitivity to changes in the $k_1$ parameter. However, near the endocardium, the hypocontractile model displayed a more sizable increase in disarray with altered $k_1$ values compared to the hypercontractile model.

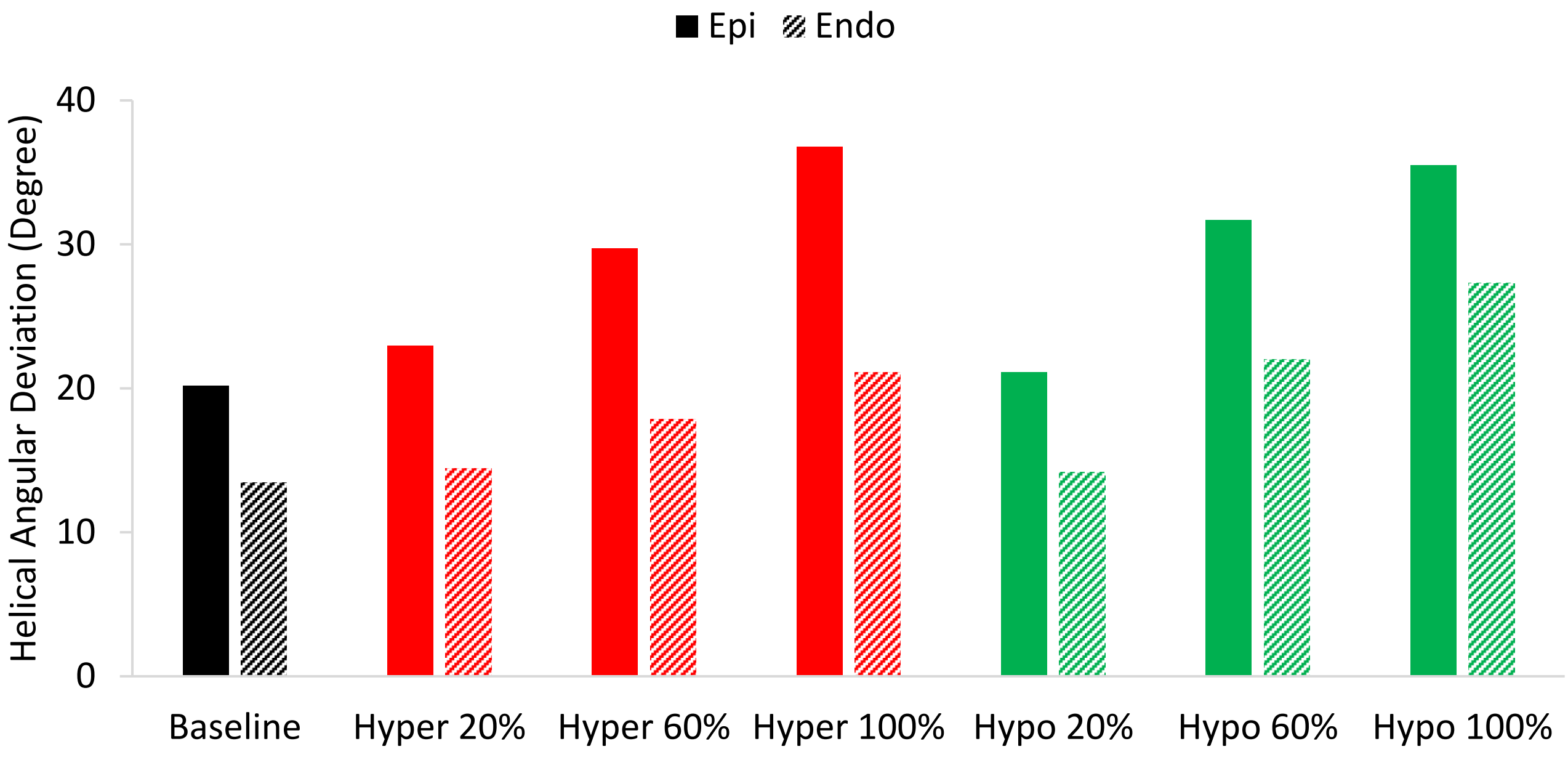

**Figure 8:** Helical angular deviations in the epicardium and endocardium across baseline, hypercontractile, and hypocontractile LV models with 20%, 60%, and 100% changes in the $k_1$ parameter. In hypocontractile models, $k_1$ is reduced compared to the baseline, while in hypercontractile models, $k_1$ is increased.

### 3.5. Comparison of systolic strain pattern

The heterogenous cell-level perturbations, along with the resultant remodeled fiber configurations, had an influence on cardiac behavior. As illustrated in Figure 9, notable regional differences emerged in the systolic strain patterns of the perturbed models compared to the baseline. In the hypocontractile and fibrous models, systolic longitudinal strain exhibited a reduction specifically within the perturbed regions. Conversely, the hypercontractile model showed an overall increase in longitudinal strain, although local decreases were also observed in some regions. In terms of the systolic circumferential strain, the fibrous model exhibiting the greatest decline in the perturbed regions, followed by the hypocontractile model. The hypercontractile model exhibited some regional variation, but the overall level of circumferential strain was comparable to baseline. Systolic radial strain displayed a transmural pattern across all cases, with variability in its alterations among the perturbed models. Notably, the fibrous model demonstrated a substantial decrease in radial strain. Overall, the fibrous model displayed the greatest decrease in all systolic strain components relative to baseline.

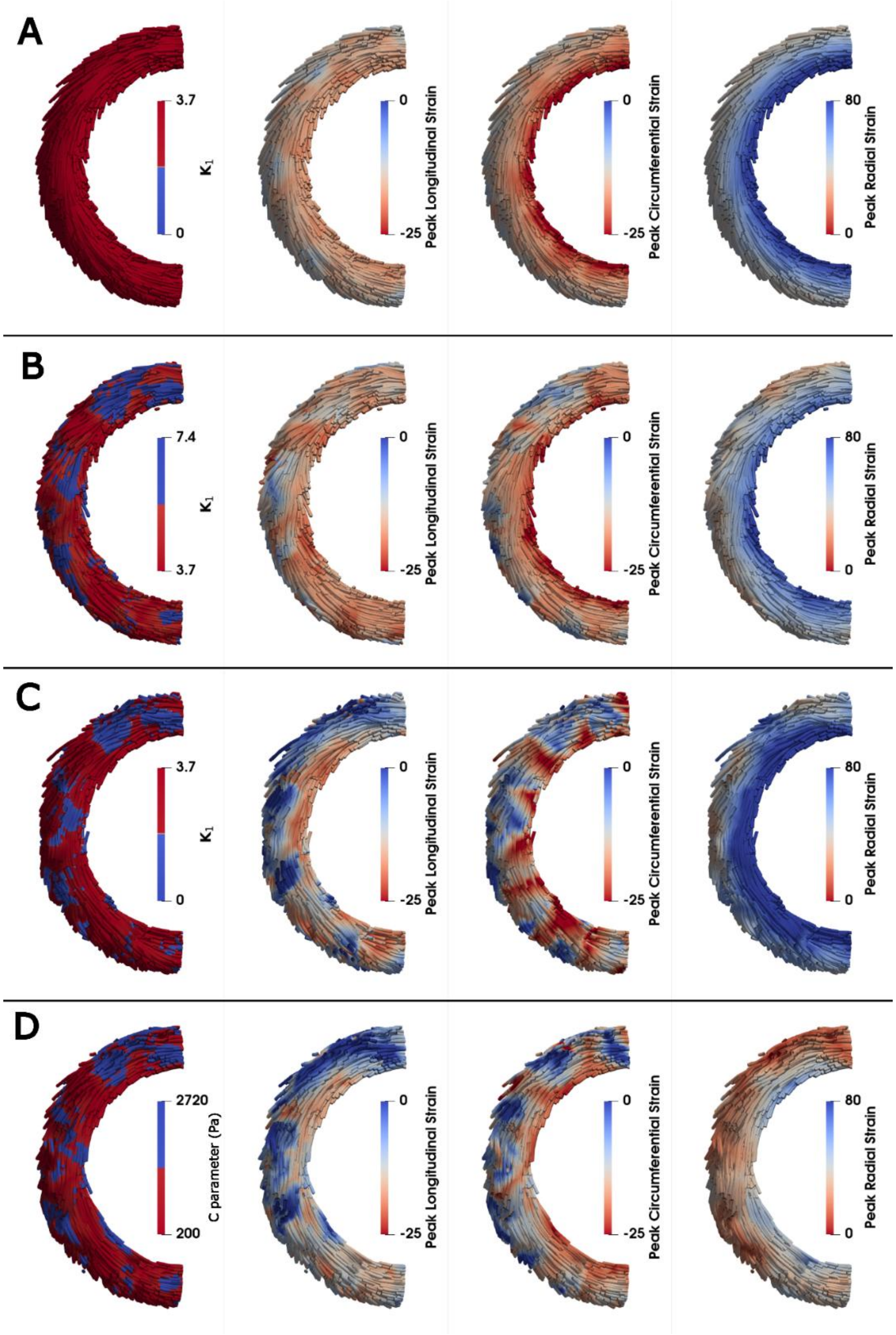

**Figure 9:** Systolic strain patterns in baseline and perturbed LV models in a midventricular short axis view reported in percent strain. **A:** Baseline, **B:** Hypercontractile, **C:** Hypocontractile, **D:** Fibrous. Note: Systolic strain was calculated by taking a multiplicative decomposition of the deformation gradient, such that the deformation gradient between end-diastole and end-systole could be used to calculate the Lagrangian strain relative to end-diastole. This is the standard approach used for reporting systolic strain [58].

### 3.6. Comparison of pumping performance

All perturbations influence the PV loop of the LV by altering its size and position (Figure 10). In the hypercontractile model, there is a minimal change in stroke volume (SV) accompanied by a leftward shift in the PV loop, resulting in a 3.5% increase in EF. Conversely, in the hypocontractile model, the PV loop shrinks and shifts to the right, leading to a 16% reduction in SV and a 23% decline in EF, along with a reduction in systolic pressure. In the fibrous model, the PV loop shrinks "inward" (i.e. ESV increases and EDV decreases) along with a reduction in systolic pressure. In the fibrous model, SV decreases by 22% and EF drops by 18%.

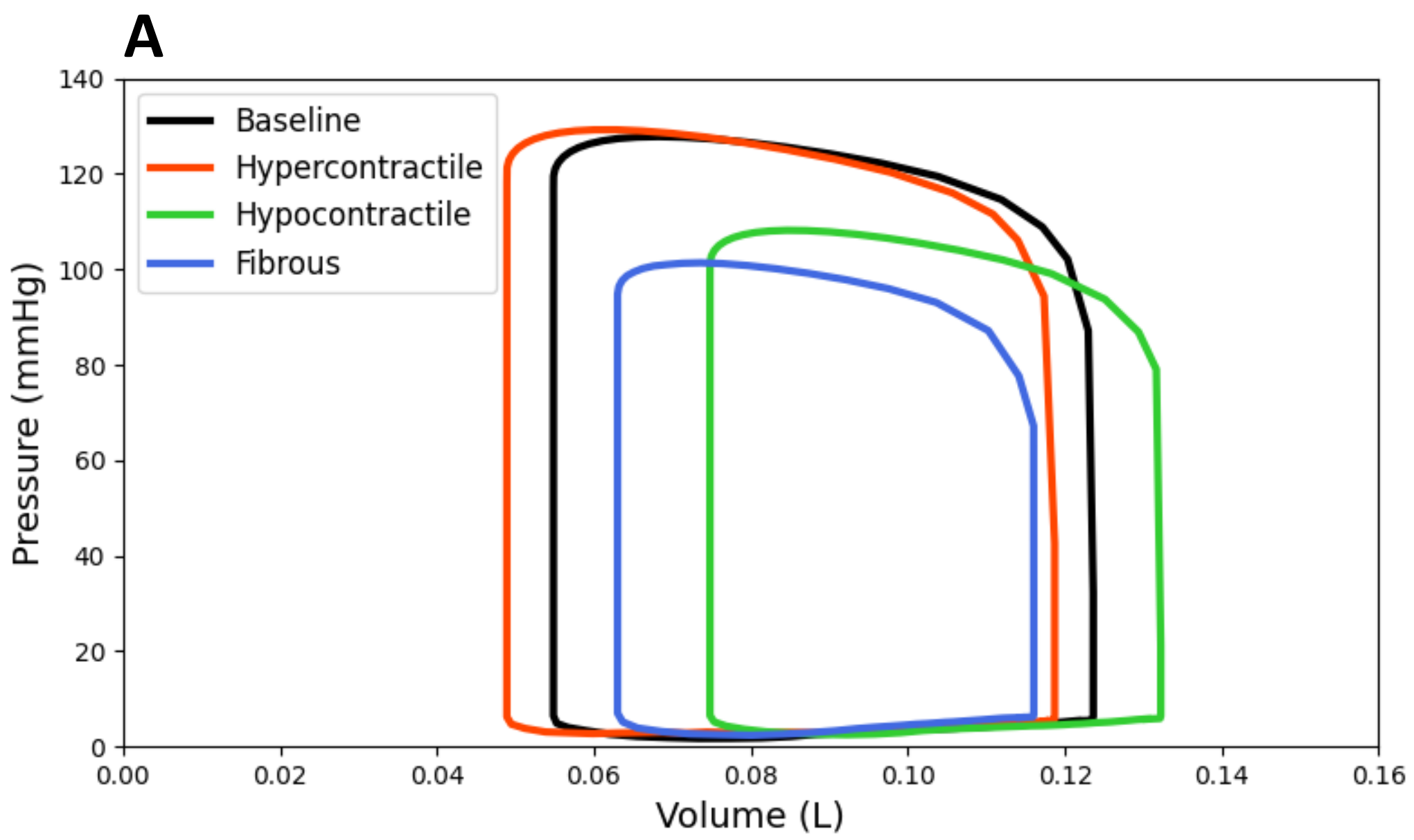

**B**

| Model | SV(L) | EF(%) |
|---|---|---|
| Baseline | 0.068 | 56 |
| Hyper | 0.069 | 58 |
| Hypo | 0.057 | 43 |
| Fibrous | 0.053 | 46 |

**Figure 10**: Cardiac pumping performance. **A:** Pressure-volume (PV) loops for baseline and perturbed LV models. **B:** Cardiac stroke volume (SV) and ejection fraction (EF) values for baseline and perturbed LV models.

### 3.7. Comparison of spatial distribution

To investigate the effects of spatial distribution on the induction of fiber disarray, we tested three sets of models with varying spatial distributions and sizes of perturbed regions (but all still affect 30% of the myocardium). In these models, the perturbed regions were progressively refined into smaller patches as showed in Figure 4 starting from Distribution 1 then further refined to Distribution 2, and finally to Distribution 3. Based on Figure 11, higher fiber disarray is observed near the epicardium compared to the endocardium in all perturbations. Near the epicardium, the highest variability was observed in the hypocontractile model, with a Coefficient of Variation (CV) of 2.5%. Meanwhile, near the endocardium, the greatest variability occurred in the hypercontractile model, with a CV of 5.2%. However, it should be noted that these relatively low CV values (both below 10%) suggest that the fiber disarray results are largely independent of the distribution of perturbed cells. Furthermore, additional analyses, where spatial distribution effects were evaluated without altering the size of the perturbed regions (see Figure S5 in Supplementary Material), confirmed the independence of fiber disarray from the distribution of perturbed cells.

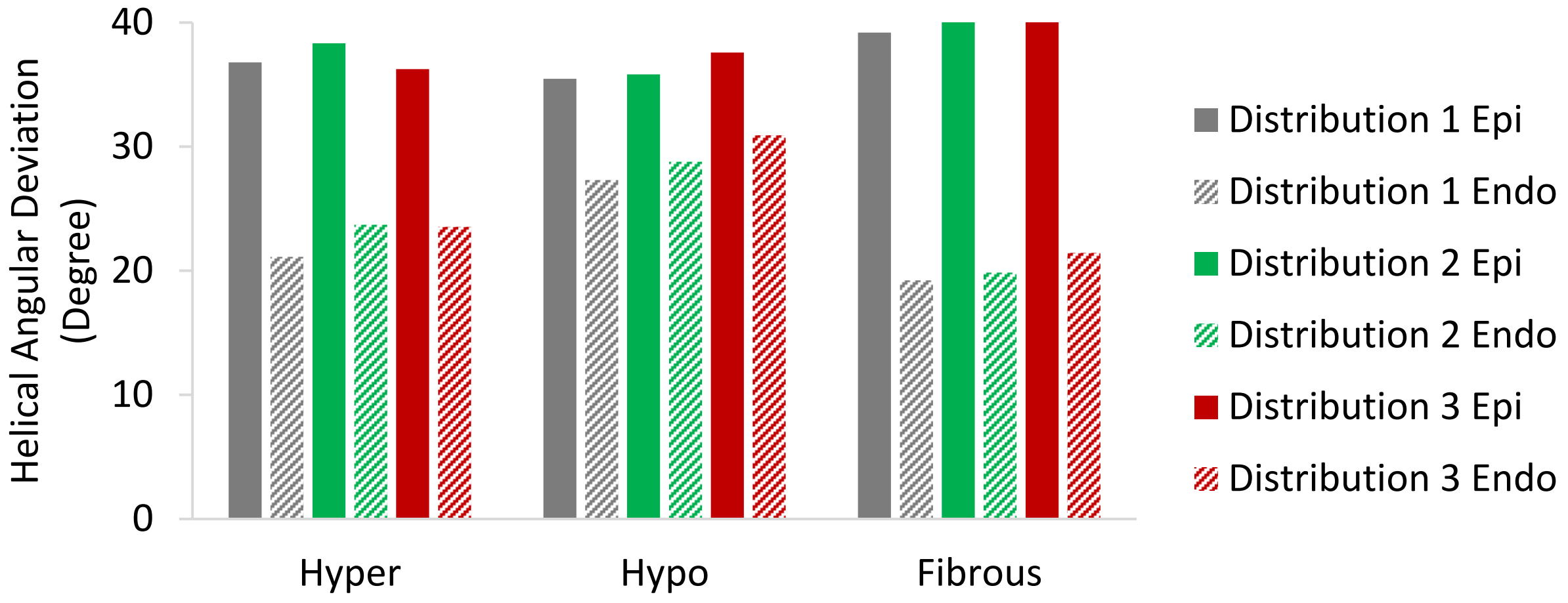

**Figure 11**: Comparison of different spatial distributions for inducing fiber disarray. Helical angular deviations in the epicardium and endocardium across perturbed LV models based on Distribution 1, Distribution 2, and Distribution 3.

## 4. Discussion:

This study investigates the development of fiber disarray in HCM and evaluates the cardiac performance of remodeled LVs. By integrating a stress-based fiber reorientation algorithm into a multiscale FE model of the LV, we explored the role of different cellular abnormalities in the development of fiber disarray. Building on our previous work [43], which demonstrated the utility of the stress-based fiber remodeling law in predicting both normal fiber configurations and pathological remodeling post myocardial infarction (MI), this study focused on how specific HCM-related abnormalities—hypercontractility, hypocontractility, and fibrosis—lead to fiber disarray and altered cardiac function. The perturbed LV models developed appreciable fiber disarray, with distinct patterns depending on the type of abnormality, offering important mechanistic insights into HCM progression.

The baseline model demonstrated smooth distributions of stress throughout the LV. However, in the perturbed cases, a significantly more heterogeneous stress distribution was observed, as shown in Figure 5. In peak systole, the active myofiber stress was primarily influenced by the contractile capacity of the myocardium, with the hypercontractile model showing regional increases in stress within the perturbed region. However, the hypocontractile and fibrotic models exhibited reduced stress in the perturbed regions with increased stress in some of the surrounding myocardium (Figure 5A). This could be due to an increase in sarcomere length within the normal myocardium at end systole (as seen in the systolic strain patterns in Figure 9), which leads to increased active stress [59]. In peak diastole, passive myofiber stress increased in the hypocontractile model (Figure 5B) due to increased end diastolic volume (i.e., the ventricle is more distended). Similar to a thick-walled pressure vessel, there is a dominant stress in the circumferential direction during diastole, particularly near the endocardium. This led to the reorientation of fibers toward the circumferential direction, as seen in Figure 6C, and contributed to the observed rise in myofiber stress. Although the fibrous model also showed increased fiber reorientation toward the circumferential direction in the endocardium, increased stiffness resulted in stress reduction in the perturbed regions (Figure 5B). In contrast, the hypercontractile model exhibited a reduction in diastolic passive stress, due to reduced end diastolic volume, and remodeling with limited reorientation toward the circumferential direction on the endocardium (remodeling was dominated by contractile stress along fibers rather than passive stress).

Interestingly, the case-specific heterogeneity in stress distribution resulted in distinct patterns of fiber disarray across the perturbed models. As illustrated in the magnified view of Figure 6A, fibers within hypercontractile regions exhibited minimal reorientation, as expected, due to the dominant contractile stress aligned with the fiber direction. However, these hypercontractile fibers exerted excessive shear stress on the surrounding normal myocardial fibers, causing them to reorient. In contrast, in hypocontractile and fibrous models, the normal myocardial fibers around the perturbed regions displayed minimal reorientation due to the dominant passive and active stress in the fiber direction. However, in the hypocontractile and fibrous regions themselves, the lack of contractility, coupled with an absence of directional dominance in passive stress (particularly in fibrous regions), resulted in substantial fiber reorientation. In this case, the elevated stress in the surrounding normal myocardium induced reorientation within the perturbed regions.

Previous studies using DT-MRI have demonstrated that a reduction in fractional anisotropy (FA) is a reliable indicator of increased myocardial fiber disarray, with lower FA values often observed in fibrotic regions. However, not all regional cases of low FA in patients with HCM can be fully corelated with fibrosis alone [4], suggesting that myocardial mechanics also plays a significant role in the severity of fiber disarray. Our models support this conclusion. Despite not incorporating any spatial preference within the myocardium when assigning perturbed regions, the angular deviation (AD) analysis (Figure 7) reveals significantly higher fiber disarray in the epicardium. This finding can be attributed to the circumferential stress (hoop stress) induced within the LV wall during loading, which has the largest magnitude near the endocardium and diminishes toward the epicardium. As a result, the endocardium experiences a dominant stress direction that aligns fibers more uniformly, reducing directional randomness and disarray. These results highlight the significant impact of mechanical stress distributions within the myocardium on fiber organization. To further investigate the relationship between impaired contractility and fiber disarray, we examined the effect of varying degrees of contractile dysfunction on fiber disarray. A nearly linear relationship was observed between fiber disarray (AD) and the $k_1$ parameter (Figure 8). This confirms the previously mentioned observation about the differences in disarray between the epicardium and endocardium, regardless of the severity of contractile imbalance. Lastly, to assess the influence of spatial distribution on our results, we modeled various spatial distributions and sizes of perturbed regions (see Figure 4B, Figure 11, and Figure S4), while preserving the 30% volume fraction of heterogeneity. Our findings indicate that fiber disarray is largely independent of the spatial distribution of perturbed cells.

The systolic strain results shown in Figure 9 align closely with previous experimental studies on normal and diseased hearts. CMR feature tracking studies have reported average values of strain using the AHA bullseye guidelines. In healthy adults, the value of longitudinal strain is approximately -16%, circumferential strain is around -20%, and radial strain is around 38% [60, 61], all of which match well with the baseline model (Figure 9A) that mimics normal cardiac function. This baseline serves as a reference for comparing the effects of different pathological conditions on myocardial strain. The perturbed models (Figure 9B-D) exhibit significant alterations in deformation patterns. The reduction in systolic strain can be attributed to different mechanisms depending on the specific cellular abnormality. In the hypocontractile case, reduced longitudinal strain reflects a decrease in contractile capacity, while in the fibrous case, the reduction in strain is linked to both reduced contractility and impaired myocardial shortening caused by increased stiffness. Notably, the hypercontractile case showed an increase in longitudinal strain in the unaffected (normal) myocardial regions. When compared to experimental data, localized increases in longitudinal strain have been reported in early-stage HCM patients [5], particularly in the non-obstructive form, which aligns with the behavior observed in our hypercontractile model. However, in more advanced stages of HCM, as documented in several studies [5, 60-65], a marked reduction in longitudinal strain is observed, consistent with the results of our fibrous model (Figure 9D). Circumferential strain followed a similar trend to longitudinal strain, with regional reductions observed in all perturbed models. With regards to radial strain, the hypocontractile and hypercontractile models showed slight deviations from the baseline case, with typical transmural gradients, while the fibrous model demonstrated a considerable reduction (also observed in CMR studies of HCM [60]), indicating compromised function of the myocardium. Overall, these results highlight the dominant role of fibrosis in altering myocardial strain patterns, particularly in its pronounced impact on longitudinal, circumferential, and radial strain.

Cardiac pumping performance was altered in almost all perturbed LV models, as demonstrated by substantial changes in the size and position of the PV loops compared to the baseline (Figure 10). In the hypercontractile model, SV remained relatively unchanged; however, the PV loop displayed a marked leftward shift (EDV 118 ml vs. 123 ml, ESV 48 ml vs. 54 ml) as reported for HCM patients [66, 67], which was also accompanied by an increase in systolic pressure. These changes are primarily due to enhanced myocardial contractility. In contrast, in the hypocontractile model, the PV loop shrank and shifted rightward, along with a noticeable reduction in systolic pressure, which reflects the reduced contractile ability of the myocardium. The fibrous model followed a similar but more severe trend than the hypocontractile model, with further PV loop shrinkage, an “inward” shift (i.e. ESV increases and EDV decreases), and a more pronounced decline in systolic pressure. This exacerbated deterioration in cardiac performance can be attributed to diminished contractile function, increased myocardial stiffness, and fiber disarray, which not only impairs systolic ejection but also diastolic filling. These observed PV loop behaviors, specifically in the hypocontractile and fibrous models, are consistent with previous findings from PV catheterization studies in patients with established HCM [66]. Overall, these results suggest that cell-level abnormalities, such as altered contractility and increased myocardial stiffness, result in significant fiber remodeling and impairments in cardiac pumping performance.

### 4.1 Limitations

This study has some limitations. First, the predictive capabilities of the MyoFE framework for fiber remodeling and resulting cardiac performance were assessed using idealized LV geometry. Future studies will incorporate more realistic geometries. Second, it should be noted that the current investigation did not consider volumetric growth, which typically occurs concurrently with HCM. This limitation will be addressed in future studies by incorporating a growth model into the modular framework and coupling it to the fiber remodeling algorithm. Another limitation of the current model is that we did not tune the time constant $\kappa$ in the reorientation law. This would require fitting to experimental data, which will be the focus of future studies. Lastly, while we modeled fibrosis by imposing the mechanical properties of fibrotic tissue, the underlying biological mechanism of fibrosis development was not implemented. Incorporating a predictive model that simulates the deposition of both diffuse fibrosis and localized replacement fibrosis as an adaptive immune response will be a focus of future research.

## 5. Conclusion

In conclusion, this study highlights the effectiveness of the multiscale cardiac modeling framework in predicting fiber remodeling within the LV. The results show that various heterogeneous perturbations in the LV models—including hypercontractility, hypocontractility, and fibrosis—disrupt the normal stress distribution within the LV muscle, leading to significant fiber remodeling. Each type of perturbation resulted in distinct patterns of fiber disarray, which offer valuable insights into the underlying mechanisms of HCM disease progression. The cumulative impact of cellular-level abnormalities and the resulting fiber disarray alter cardiac performance and functional characteristics, with unique patterns emerging for each perturbation. This comparative analysis provides a deeper understanding of structural and functional cardiac anomalies, potentially aiding in the identification of specific cellular-level causes linked to cardiac dysfunction.

## Acknowledgments:

Support for this research was provided by National Institutes of Health grant R01 HL163977 and National Science Foundation grant IIS-2406028.

## Author Contributions:

JFW, LCL, and KSC conceived the study design. MM developed the finite element model and implemented the fiber reorientation algorithm. MM ran all of the simulations and analyzed the results. JFW, LCL, and KSC assisted in the interpretation of the results. All authors contributed to the review and finalization of the manuscript.

## Competing Interests:

The author(s) declare no competing interests.

## Data Availability Statement:

The MyoFE code is maintained in a github repository. The code will be made available to other researchers upon request to the corresponding author.

# Supplementary Materials

## Supplementary tables

Table S1. Model parameters for baseline simulation.

| Component | Parameter | Value | Units |
|---|---|---|---|
| Lumped parameter model of systemic circulation | HR | 65 (bpm) | (bpm) |
| | $V_{total}$ | 4.5 (liters) | (liters) |
| | $R_{aorta}$ | 25.0 | (mmHg $L^{-1}$ s) |
| | $R_{arteries}$ | 25.0 | (mmHg $L^{-1}$ s) |
| | $R_{arterioles}$ | 787.0 | (mmHg $L^{-1}$ s) |
| | $R_{capillaries}$ | 350.0 | (mmHg $L^{-1}$ s) |
| | $R_{venules}$ | 50.0 | (mmHg $L^{-1}$ s) |
| | $R_{veins}$ | 50.0 | (mmHg $L^{-1}$ s) |
| | $R_{LV}$ | 10.0 | (mmHg $L^{-1}$ s) |
| | $C_{aorta}$ | 0.00035 | ($mmHg^{-1}$ L $s^{-1}$) |
| | $C_{arteries}$ | 0.0008 | ($mmHg^{-1}$ L $s^{-1}$) |
| | $C_{arterioles}$ | 0.001 | ($mmHg^{-1}$ L $s^{-1}$) |
| | $C_{capillaries}$ | 0.01 | ($mmHg^{-1}$ L $s^{-1}$) |
| | $C_{venules}$ | 0.03 | ($mmHg^{-1}$ L $s^{-1}$) |
| | $C_{veins}$ | 0.07 | ($mmHg^{-1}$ L $s^{-1}$) |
| | $V_{aorta,slack}$ | 0.3 | (liters) |
| | $V_{arteries,slack}$ | 0.3 | (liters) |
| | $V_{arterioles,slack}$ | 0.1 | (liters) |
| | $V_{capillaries,slack}$ | 0.25 | (liters) |
| | $V_{venules,slack}$ | 0.5 | (liters) |
| | $V_{veins,slack}$ | 2.0 | (liters) |
| Myocardium passive properties | C | 200.0 | (Pa) |
| | $b_{ff}$ | 8.0 | (Unitless) |
| | $b_{xx}$ | 3.58 | (Unitless) |
| | $b_{fs}$ | 1.63 | (Unitless) |
| | $C_1$ | 250.0 | (Pa) |
| | $C_2$ | 15.0 | (Unitless) |
| MyoSim model of contraction | $k_{on}$ | $1.7\times10^{8}$ | ($M^{-1}$ $s^{-1}$) |
| | $k_{off}$ | 200 | ($s^{-1}$) |
| | $k_{coop}$ | 5 | (Unitless) |
| | $k_1$ | 3.7 | ($s^{-1}$) |
| | $k_2$ | 200 | ($s^{-1}$) |
| | $k_3$ | 100 | ($s^{-1}$ $nm^{-1}$) |
| | $k_{4,0}$ | 80 | ($s^{-1}$) |
| | $k_{4,1}$ | 1.5 | ($s^{-1}$ $nm^{-4}$) |
| | $k_{cb}$ | 0.001 | (pN $nm^{-1}$) |
| | $x_{ps}$ | 5 | (nm) |
| | $N_0$ | $6.9\times10^{16}$ | ($m^{-2}$) |
| | $k_B$ | $1.38\times10^{-23}$ | ($JK^{-1}$) |
| | T | 310 | (K) |
| Two compartment model of $Ca^{2+}$ transient | $k_{SERCA}$ | 7.866 | ($M^{-1}$ $s^{-1}$) |
| | $k_{act}$ | $8.06\times10^{-2}$ | ($M^{-1}$ $s^{-1}$) |
| | $k_{leak}$ | $6\times10^{-4}$ | ($M^{-1}$ $s^{-1}$) |
| Fiber reorientation law | $\kappa$ | 4000 | (ms) |

**Supplementary figures:**

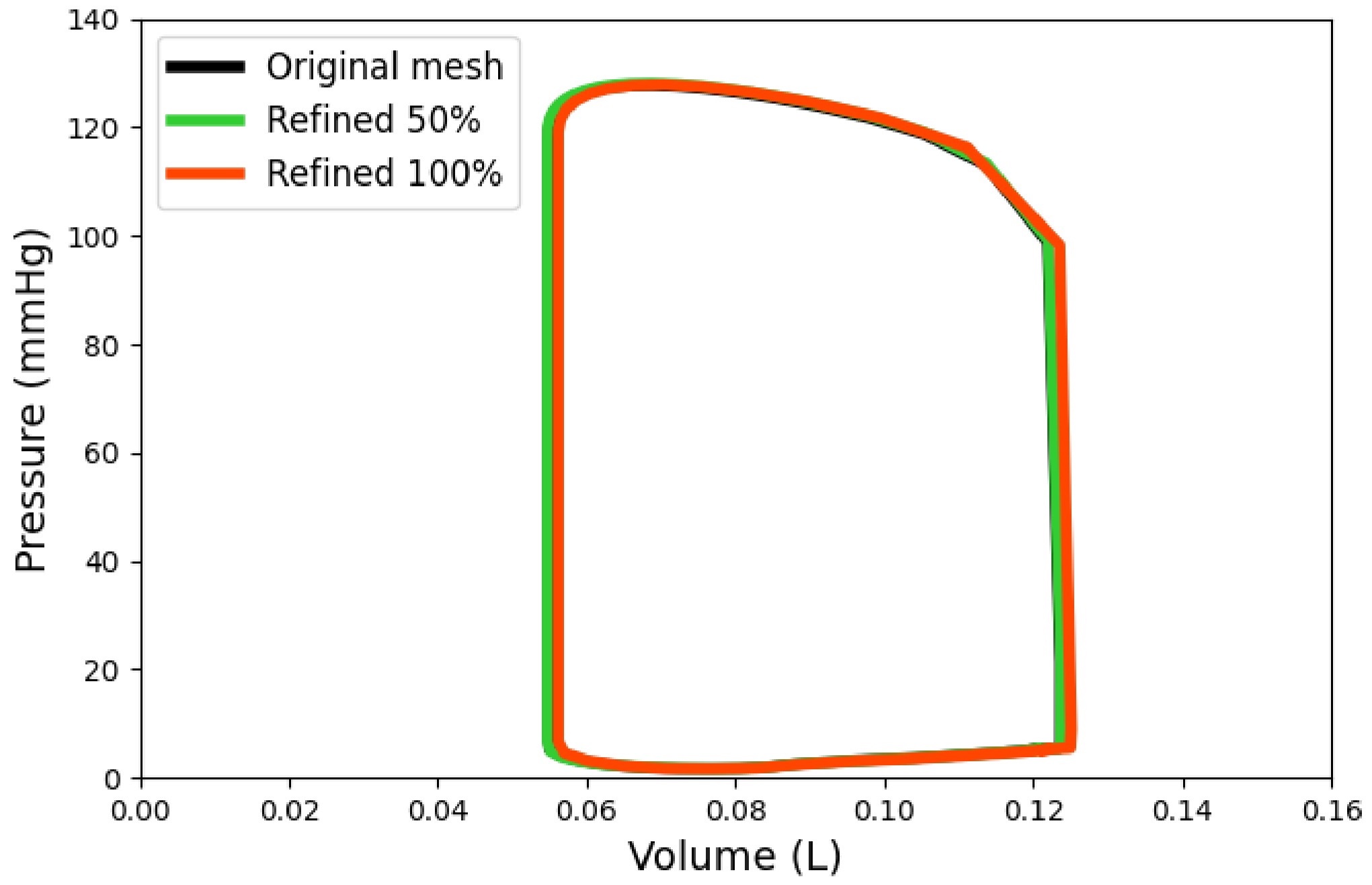

**Figure S1:** Comparison of Pressure-Volume (PV) loops in LV Models with different levels of mesh refinement. The original mesh contains ~1250 quadratic tetrahedral elements. Note that there is minimal variation between the loop plots.

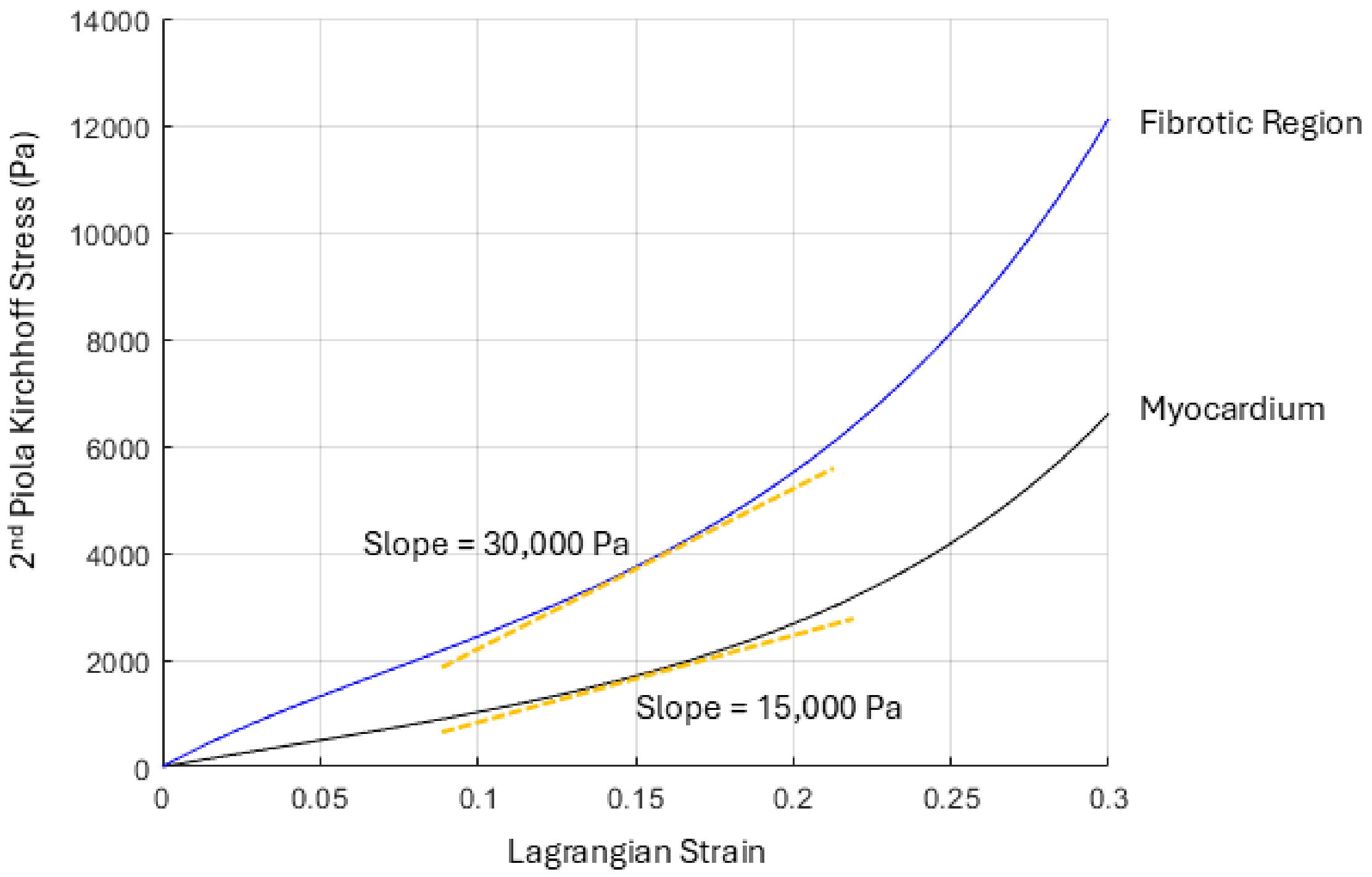

**Figure S2:** Stress-strain curves of the myocardium and fibrotic regions. These represent the mechanical response along the fiber direction. Stiffness was calculated by taking the slope of the stress-strain curves at a strain value of 0.15 (i.e., 15% strain).

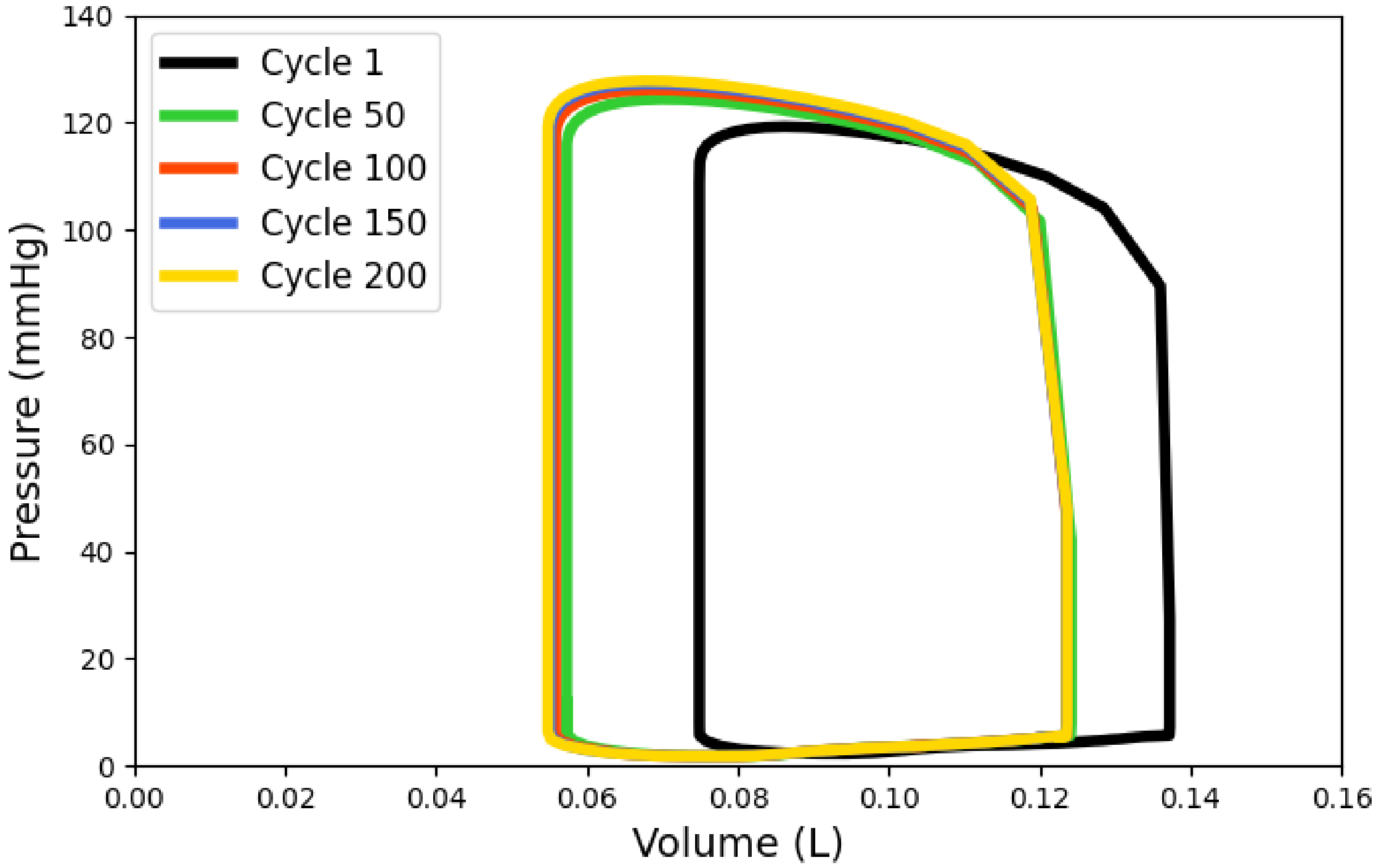

**Figure S3:** Evolution and convergence of Pressure-Volume (PV) loops in LV model. PV loops are plotted for the 1st, 50th, 100th, 150th and 200th cardiac cycles with fiber reorientation.

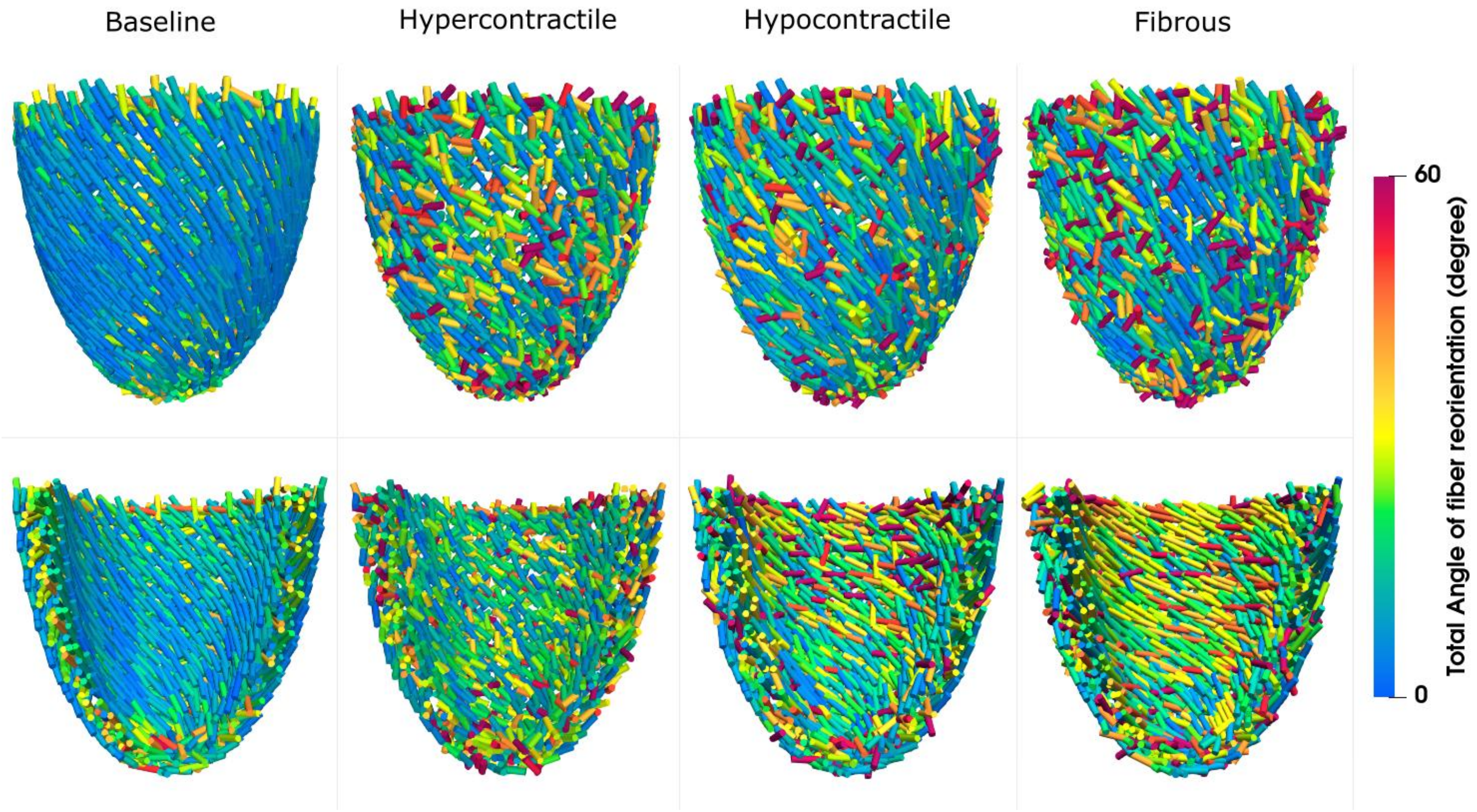

**Figure S4:** Comparison of fiber disarray between baseline and perturbed LV models based on the total angle of fiber reorientation. Top: Epicardial side of the myocardium. Bottom: Endocardial side of the myocardium.

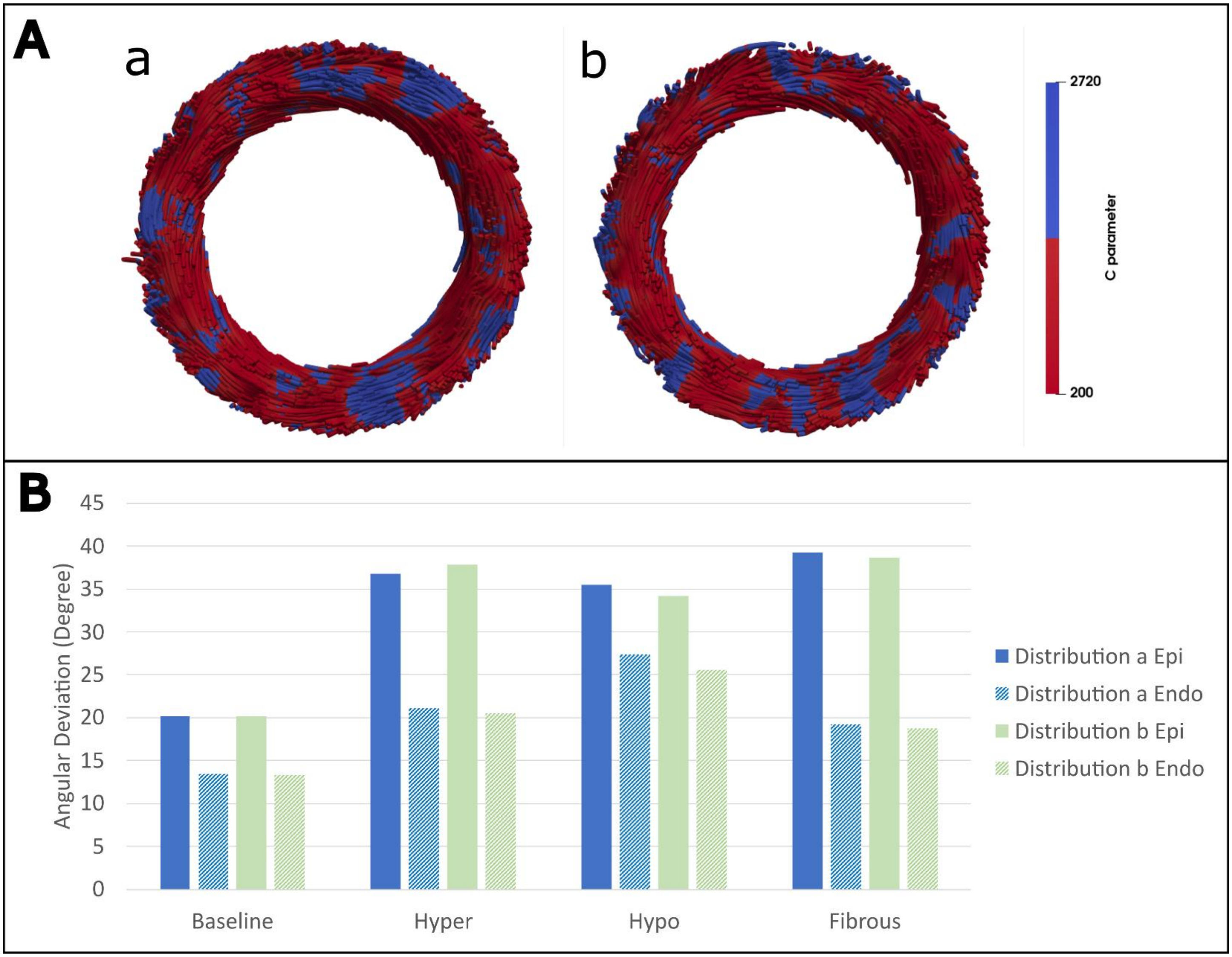

**Figure S5:** Spatial variability without variation in size of perturbed regions, i.e., the heterogenous perturbed tissue consisted of 30% of the LV myocardium and the minimum size of a perturbed region was 3 mm. **A:** Short axis view of mid ventricle with perturbed distribution of a and b. **B:** Induced disarray from distribution a and b in baseline and perturbed LV models.